# Driving factors of auditory category learning success

Nan Wang [a], Gangyi Feng [a,b,*]

a Department of Linguistics and Modern Languages, The Chinese University of Hong Kong, Shatin, N.T., Hong Kong SAR, China

b Brain and Mind Institute, The Chinese University of Hong Kong, Shatin, N.T., Hong Kong SAR, China

* Corresponding author

Gangyi Feng

Brain and Mind Institute

Department of Linguistics and Modern Languages

The Chinese University of Hong Kong

Shatin, N.T., Hong Kong SAR, China

g.feng@cuhk.edu.hk

**Funding**

The work described in this paper was supported by grants from the Research Grants Council of the Hong Kong Special Administrative Region, China (Project No.: 14614221, 14612923, 14621424, and C4001-23Y to Gangyi Feng) and the National Natural Science Foundation of China (Project No. 32322090 to Gangyi Feng).

**Abstract**

Our brain learns to update its mental model of the environment by abstracting sensory experiences for adaptation and survival. Learning to categorize sounds is one essential abstracting process for high-level human cognition, such as speech perception, but it is also challenging due to the variable nature of auditory signals and their dynamic contexts. To overcome these learning challenges and enhance learner performance, it is essential to identify the impact of learning-related factors in developing better training protocols. Here, we conducted an extensive meta-analysis of auditory category learning studies, including a total of 111 experiments and 4,521 participants, and examined to what extent three hidden factors (i.e., variability, intensity, and engagement) derived from 12 experimental variables contributed to learning success (i.e., effect sizes). Variables related to intensity and training variability outweigh others in predicting learning effect size. Activation likelihood estimation (ALE) meta-analysis of the neuroimaging studies revealed training-induced systematic changes in the frontotemporal-parietal networks. Increased brain activities in speech and motor-related auditory-frontotemporal regions and decreased activities in cuneus and precuneus areas are associated with increased learning effect sizes. These findings not only enhance our understanding of the driving forces behind auditory category learning success, along with its neural changes, but also guide researchers and practitioners in designing more effective training protocols that consider the three key aspects of learning to facilitate learner success.

## 1. INTRODUCTION

Learning to abstract sensory experiences into categories is a fundamental human ability essential for adaptation and survival (Grinband et al., 2006; Pedrosa et al., 2023; E. E. Smith & Grossman, 2008; J. D. Smith et al., 2012). A key aspect of this ability is the categorization of auditory stimuli, which enables us to distinguish critical environmental sounds (e.g., differentiating between a car horn and a bicycle bell) and to perceive and produce speech for effective communication (e.g., discerning whether a friend said “right” or “light”) (Goldstone & Hendrickson, 2010; Hickok & Poeppel, 2007). However, learning to categorize sounds poses significant challenges due to the acoustically variable and temporally transient nature of auditory signals and their surrounding dynamic contexts (Obasih et al., 2023). This learning process involves not only repeated exposure to varied exemplars but also the practice of categorization as we interact with our changing environment, which enables us to recognize sounds and their categories efficiently in new contexts. Understanding the stimulus and contextual factors contributing to successful auditory category learning is crucial for developing effective training paradigms that facilitate neural changes and enhance learning outcomes.

Previous research has explored various training paradigms and variables to examine how they shape brain functions and improve auditory category learning, such as variability in stimulus and presentation procedure, intensity of exposure, and learning strategies (Gabay et al., 2015; Gan et al., 2023; Maddox et al., 2008a; Thorin et al., 2018; Worthy et al., 2013). However, many studies have focused on specific and limited sets of variables, and there is a lack of a systematic framework to compare the relative contributions of different factors to behavioral outcomes and the underlying neural changes. Consequently, findings across studies have been mixed and sometimes contradictory, making it challenging to determine the most effective sets of variables to enhance learning outcomes. Thus, it is essential to identify the key factors that contribute to successful auditory category learning and to understand the related brain changes using a systematic meta-analytical approach that synthesizes extensive behavioral and neuroimaging studies.

Neural systems underlying the perception of auditory categories involve distributed brain networks. Extensive studies were conducted in the field of speech. Especially, over the past decades, significant progress has been made in elucidating the cortical mechanisms underlying speech perception, particularly the mapping of auditory signals onto meanings via the ventral pathway and onto articulatory units through the dorsal pathway (Hickok & Poeppel, 2000, 2007;

Saur et al., 2008). While the dual pathway models serve as insightful frameworks for understanding the neural processing and representation of well-learned auditory stimuli, especially native speech units, it remains largely unclear how the human brain reorganizes to acquire novel auditory category knowledge through learning. Furthermore, extending these models to general auditory category representation and learning requires further investigation, since non-speech sounds may recruit both shared and unique mechanisms. Previous studies show that while speech categories depend on language and speech systems, they also involve domain-general auditory pathways, both of which are shaped by experience (Holt & Lotto, 2010; Kuhl, 2004a; Wong et al., 2004; Zatorre et al., 2002).

During the learning of auditory categories, previous studies have identified multiple neural networks involved, including the auditory temporal cortex, frontoparietal cortices, basal ganglia, hippocampus, and motor-related areas (Bartolotti et al., 2017; Desai et al., 2008; Feng et al., 2019; Karuza et al., 2014; S.-J. Lim et al., 2014; Myers, 2014; Myers & Swan, 2012; Wang et al., 2003a; Yi et al., 2016). Notably, models within the Multiple Learning Systems (MLS) framework, including the Dual Learning Systems (DLS) model, propose that two distinct neural circuits are involved, depending on the category structures and the learning strategies employed by learners (Ashby & Maddox, 2005; Chandrasekaran et al., 2014; Maddox et al., 2013; McMurray, 2023; Roark & Chandrasekaran, 2023; Roark & Holt, 2019). The rule-based learning stream involves the superior temporal gyrus (STG), prefrontal cortex, hippocampus, and head of the caudate nucleus (Doeller et al., 2006; Filoteo et al., 2005; Helie et al., 2010) while the reward-based learning stream involves the STG, precentral gyrus, and ventral striatum (e.g., putamen) (Baumeister et al., 2019; Cox & Witten, 2019; Feng et al., 2021; Galvan et al., 2005; Guo et al., 2013; Heo et al., 2021; Schönberg et al., 2007).

While these findings and models offer valuable insights into the neural substrates of auditory category learning, it remains unclear how these regions and networks reorganize during and after training to support effective learning and categorization. Training has often been associated with enhanced brain activity in auditory and speech-related regions, indicating improved processing and representation of features relevant to categorization (Callan et al., 2003; Feng et al., 2019; Wang et al., 2003b). Conversely, decreased brain activity may indicate increased neural efficiency and reduced reliance on ancillary processes as learning progresses (James & Gauthier, 2003; E. E. Smith & Grossman, 2008; Yotsumoto et al., 2008). However, these decreases are less documented, likely due to limitations in sample sizes for individual neuroimaging studies (Grady et al., 2020; Müller et al., 2017) and differences in how training variables were manipulated across studies (Poldrack et al., 2017). Thus, meta-analyses involving neuroimaging studies are essential for identifying reliable patterns of neural changes and understanding the factors that contribute to auditory category learning success across both speech and non-speech domains.

In this study, we focus on examining three sets of training-related variables: variability, intensity, and engagement, which have garnered increasing research attention in recent years and may play important roles in shaping learning success and neural changes. Variability in training is a prominent area of investigation in auditory category learning research, often hypothesized to induce neural changes, promote effective learning, and enhance performance in generalization to new contexts. However, the outcome effects remain controversial (K. G. Estes & Lew-Williams, 2015; Raviv et al., 2022). Training variability involves manipulating the diversity of stimuli and contexts by changing talkers (e.g., using different speakers), contexts (e.g., varying phonetic contexts), modalities (e.g., auditory versus audiovisual presentation), or presentation sequences (e.g., interleaved versus blocked) (Carvalho & Goldstone, 2014, 2015; Gabay et al., 2015; S. Lim & Holt, 2011; Schorn & Knowlton, 2021). High-variability training exposes learners to a wide range of exemplars, potentially facilitating the abstraction of category-relevant features and leading to better performance when categorizing unseen items (Gabay et al., 2015). However, empirical evidence on the benefits of high variability training is mixed (Brekelmans et al., 2022). Some studies suggest that high variability may hinder learners with weaker perceptual abilities while benefiting those with stronger abilities (Perrachione et al., 2011; Sinkeviciute et al., 2019). Others find that the benefits of variability depend on the type of stimulus, enhancing identification in familiar speech materials but not in novel non-speech sounds (Sadakata & McQueen, 2013), or the stage of learning, with high variability hindering early acquisition but facilitating better generalization to novel sounds at later stages (Braithwaite & Goldstone, 2015; Raviv et al., 2022). These inconsistencies suggest that the effects of variability could be context-dependent and may interact with individuals' characteristics and other training factors, requiring further examination.

Training intensity is another crucial dimension that contributes to learning success. Training intensity refers to the amount of exposure and practice and is manipulated through variables such as the number of trials, duration, and training frequency. Increased training intensity is generally associated with more significant learning gains (D. Roth, 2005; Drullman & Bronkhorst, 2004), but it may also lead to cognitive fatigue or overlearning, potentially diminishing efficiency (Joiner & Smith, 2008; Molloy et al., 2012). Extended training duration also increases the opportunity for the critical learning process of consolidation to occur. The importance of consolidation, especially sleep-dependent consolidation, has been highlighted in stabilizing and enhancing learning (Plihal & Born, 1997; Tucker et al., 2006). The manipulation of training intensity is closely related to learning progress and the stage of learning. High-intensity early training may facilitate novice learners' transition to subsequent learning stages (Rohrer & Taylor, 2006), potentially enabling them to benefit more rapidly from other factors such as high training variability, motivation, and engagement. These dynamic interactions between factors require further investigation.

Learning engagement is another putative hidden factor contributing to auditory category learning success. During training, engagement can be implemented through variables such as feedback, learning task (e.g., active versus passive tasks), training instruction, and motivation (e.g., monetary rewards) (Deci et al., 1999; S. Lim & Holt, 2011; Obasih et al., 2023). These variables can induce different learning strategies during the learning process, thereby affecting outcomes (Fredricks et al., 2004; Trowler, 2010). Providing feedback is generally considered beneficial for learning, as it informs learners of their performance (Hattie & Timperley, 2007). However, studies in auditory category learning have yielded mixed findings. For example, McCandliss et al. (2002) found no significant difference between feedback and no-feedback conditions in phoneme training (categorizing /l/ and /r/) for Japanese speakers when an adaptive training procedure was used. The effectiveness of feedback can depend on both its timing and the nature of the learning task. Immediate feedback is beneficial for rule-based learning but may not help with learning information-integration categories (Maddox et al., 2008b; Worthy et al., 2013). Similarly, the roles of explicit instruction in auditory category learning are multifaceted. While explicit instruction and clear descriptions of category distinctions can enhance learning in some situations (Nishi & Kewley, 2007), implicit training approaches that do not emphasize phonetic differences can also lead to similar or even better outcomes (S.-J. Lim et al., 2019; Vlahou et al., 2011). The debate over the effectiveness of active versus passive learning tasks continues, with some findings in favor of active engagement (Hammill et al., 2022) and others pointing to the benefits of implicit exposure in learning (Kaufman et al., 2010).

In this study, we examined the impact of the three latent factors on auditory category learning success by integrating behavioral and neuroimaging data from the existing literature. We conducted a comprehensive meta-analysis that included 111 experiments with 4,521 participants, systematically examining 12 training-related variables. These variables are categorized into the three factors: variability, intensity, and engagement. Utilizing confirmatory factor analysis (CFA), general linear modeling (GLM), and activation likelihood estimation (ALE), we validated the variable-factor division and assessed the contributions of these training variables to the learning effect size and neural changes associated with auditory category learning. We quantified the contributions of these factors and identified consistent patterns of changes in brain activation associated with auditory category learning across studies.

## 2. METHODS

### 2.1 Literature search

Literature searches for the meta-analysis of behavioral and neuroimaging studies are conducted separately. This process identified 111 studies, including 104 for the behavioral meta-analysis and 29 neuroimaging studies for the neuroimaging meta-analyses, with 22 studies shared between the two analyses  (see the study list in Supplementary Table S1).

For the behavioral meta-analysis, we utilized a combination of two terms. The first set of terms included "auditory category," "sound category," or "speech category," while the second term focused on learning and training, specifically "learning" or "training." The search was conducted in Web of Science and PubMed from January 2000 to March 2024, resulting in an initial yield of 5,959 papers.

For neuroimaging studies, we conducted a literature search through Web of Science and PubMed, focusing on papers published between January 2000 and March 2024. Three sets of search terms, which were combined using the logical operator, AND, were run in the database: (1) auditory-related terms (i.e., "speech" OR "phonology" OR "perceptual" OR "tone" OR "speech category" OR "lexical tone" OR "vowel" OR "pitch" OR "intonation" OR "prosody" OR "phoneme" OR "sound category" OR "music" OR "melody" ), (2) learning-related terms (i.e., "learning" OR "training" OR "acquisition" OR "discrimination" OR "practice"), and (3) neuroimaging terms (i.e., "fMRI" OR "functional magnetic resonance imaging" OR "functional neuroimaging"). This initial search yielded 7178 papers.

The screening criteria applied to both neuroimaging and behavioral studies were as follows: the studies had to be empirical research, published in English, and involve at least one group of healthy adults, excluding participants with diagnosed psychiatric or neurological disorders. Studies also needed to explicitly report effect sizes or sufficient data, such as sample size, means, and standard deviations, for effect size calculation. Also, studies should involve at least an auditory or speech category training session. Studies that focused solely on speech or auditory perception without any learning component were excluded. Only studies using fMRI as the imaging modality were included, and the studies had to report peak coordinates in Talairach or MNI (Montreal Neurological Institute) space.

Because the behavioral and neuroimaging literatures differ in size and how tasks are described, we used slightly different search strategies for each type. Both searches started with the same broad descriptors ("auditory category," "sound category," or "speech category" combined with "learning" or "training"), which cover most behavioral studies. The neuroimaging literature tends to be smaller and more likely to describe procedures tailored for fMRI settings. To maximize coverage of non-speech auditory learning studies, we added additional terms, such as "music," "melody," and "pitch," to the neuroimaging search. However, including these initial terms did not change the final set of eligible fMRI studies, as all records were screened using the same PRISMA criteria (Figure 1) and only those explicitly studying auditory category learning were included. The distribution of speech and non-speech studies in both behavioral and neuroimaging datasets can be found in Supplementary Table S1.

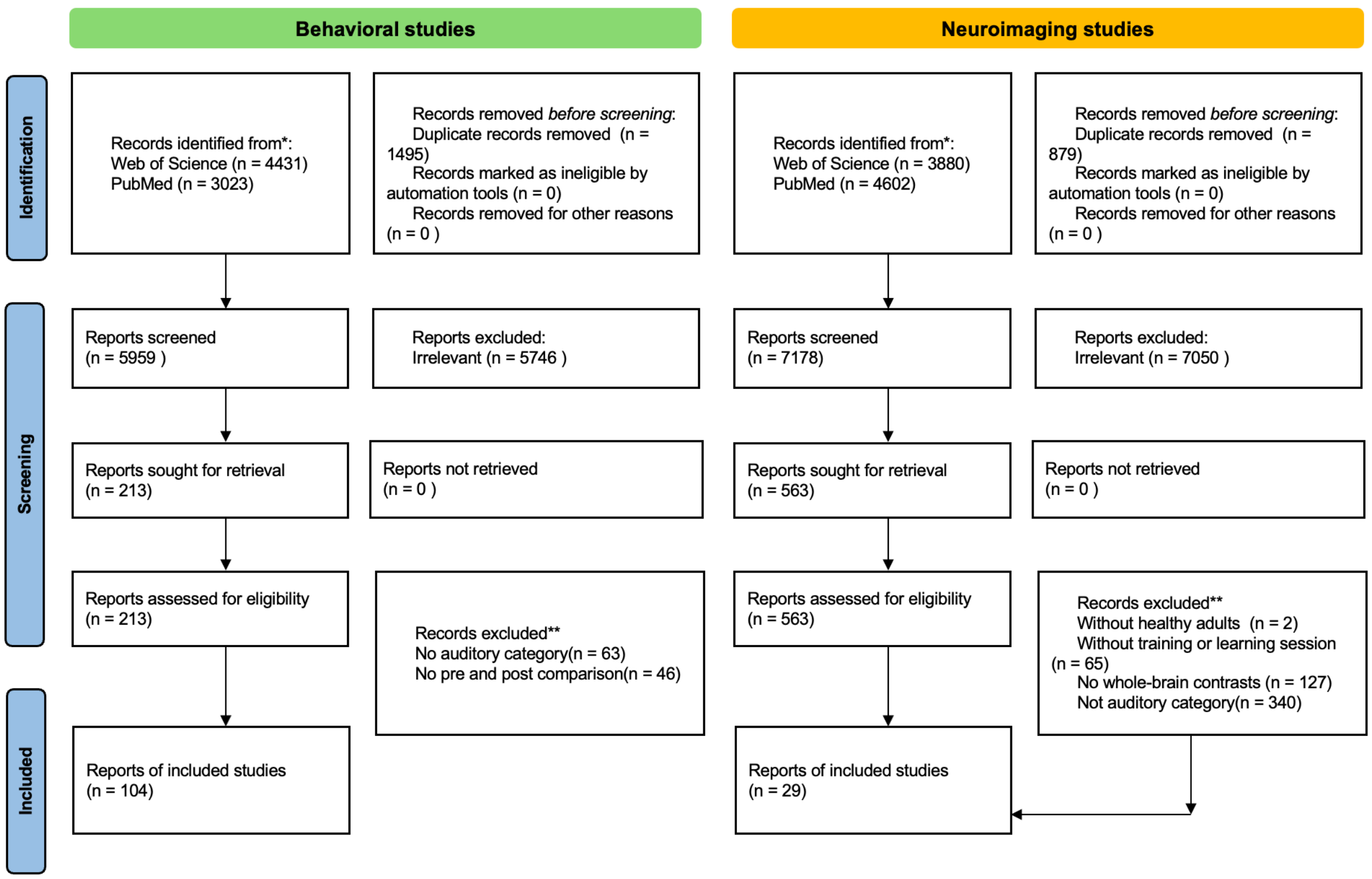

**Figure 1**. Flow diagram for identification of behavioral and neuroimaging studies related to auditory category learning based on the protocol of Preferred Reporting Items for Systematic Reviews and Meta-Analysis (PRISMA).

## 2.2 Factors and variables of interest

We selected variables to represent each of the three training factors (i.e., variability, intensity, and engagement) to examine the extent to which these variables and factors contribute to the effect sizes of learning. Variability-related variables include talker variability, context variability, modality variability, sound type, presentation variability, and exemplar variability. Engagement-related variables consist of learning tasks, feedback, and instruction. Increased engagement may arise from active learning tasks, timely feedback, and explicit instructions. Intensity-related variables encompass the number of trials, total training duration (in minutes), and the number of training days, which collectively reflect the overall training intensity.

We grouped the coded variables into three latent factors (i.e., variability, intensity, and engagement), based on theoretical considerations from the auditory category learning literature and validated with confirmatory factor analysis (Figure 2). Variability-related variables capture differences in the diversity of training input. This includes talker variability (single versus multiple speakers), context variability (number of phonetic contexts), modality variability (single [auditory] versus multiple modalities [audiovisual]), and presentation variability (blocked versus interleaved). We also included sound type (speech versus non-speech) under variability factor because prior research emphasizes that learners must overcome the substantial variability in speech input to form robust categories (Chandrasekaran et al., 2014; Holt & Lotto, 2010; McMurray & Aslin, 2005), whereas non-speech sounds in our included studies, like pure tones or artificial sounds, generally contain less natural variations (Zatorre et al., 2002). Finally, we included exemplar variability by quantifying exemplar repetitions, where no repetitions indicate high exemplar variability. Intensity-related variables (i.e., training duration, number of trials, training days) reflect the overall amount of training, while engagement-related variables (i.e., feedback, instruction, and task type) measure the degree of active involvement or explicit guidance during training. This conceptual framework served as the basis for our analyses and was empirically validated through confirmatory factor analysis.

### 2.3 Variable coding procedure

*Talker variability* (talker var): This variable is coded by the number of speakers used in each experiment. The inclusion of more speakers in training indicates greater acoustic variations. These acoustic variations are often caused by differences in the vocal tract morphology and physiology of various speakers (Kartushina & Martin, 2019; Lammert et al., 2013).

*Context variability* (context var): This variable refers to the variations in phonetic contexts surrounding a target phoneme to be learned. For example, when learning to distinguish /ʌ/ from other vowels, the two words /hVt/ and /bVt/ represent two distinct phonetic contexts. We first counted the number of phonetic contexts participants encountered during learning, which ranged from 0 (no variability in phonetic context) to 132 (132 minimal pairs of real words containing /l/ or /r/ ) across studies (Mean = 9.78, SD = 17.92), and then categorized them into three levels: low (0), medium (1-10), and high (more than 10). The threshold is set to balance the number of studies across different levels, preventing the overrepresentation of any single level. Note that the operationalization of context and talker variability is specific to speech studies and likely underestimates similar sources of variability in non-speech studies (see Supplementary Table S1 for more details about the stimulus type used in each study).

*Modality variability* (modality var): It refers to the number of modalities through which learning occurs. It is coded as 0 when the training includes only the auditory modality and as 1 when multimodal presentation is utilized (e.g., Hardison, 2003; Kartushina & Martin, 2019). In the context of auditory category learning, multimodality involves: a) providing auditory, visual, or oral feedback; b) training in one modality while testing in another.

*Sound type*: It refers to whether the training materials are speech or non-speech. It was coded as 0 (non-speech) if the stimuli used were artificially generated and as 1 (speech) if they were recorded from humans.

*Presentation variability* (present var): It refers to whether the presentation of stimuli is interleaved or blocked during training. In an interleaved presentation, auditory items are presented randomly (e.g., /r/, /l/, /s/), whereas in a blocked presentation, participants listen to phonemes from one category exclusively before moving on to another (e.g., /r/, /r/, /r/…, /l/, /l/, /l/…). Interleaved presentation presumably generates more variety in learning items than blocked presentations. It is coded as 1 for interleaved and 0 for blocked.

*Exemplar variability* (exem var): This variable measures the diversity of exemplars used during training. We operationalized this variable by determining whether the training exemplars were repeated. It was scored as 0 if sound materials were repeated (likely encountering fewer exemplars, indicating lower variability) and 1 if no repetitions occurred (likely encountering more exemplars, indicating higher variability).

*Instruction*: This variable indicates whether participants received explicit guidance to support category learning. It was coded as 1 when studies provided explicit learning instructions and 0 when no such instructions were given. Explicit instructions refer to detailed verbal or written explanations designed to help distinguish between auditory patterns, for example, a description of how pitch contours relate to different lexical tones (Nishi & Kewley, 2007). This variable is related to the explicitness of learning, a concept widely discussed in the literature.

*Feedback*: It is categorized into three levels: no feedback, minimal feedback, and full feedback. No feedback indicates the absence of any evaluative input during training. Minimal feedback consists of simple correctness feedback (e.g., indicating whether a response is correct or incorrect). Full feedback involves providing detailed differences between two or more sound categories. It is coded as 0 for no feedback, 1 for minimal feedback, and 2 for full feedback.

*Learning task*: This variable reflects whether participants engaged in an active or passive learning task. Passive learning involved simply being exposed to auditory stimuli and their labels (e.g., listening while viewing category labels), while active learning involved explicit task involvement, such as categorization or production. It was coded as 1 for active learning and 0 for passive exposure. This procedure-based variable, along with the Instruction variable, serves as a practical proxy of whether learning was more likely to occur through explicit learning or via more implicit, exposure-driven learning.

*Training duration* (duration): Total time allocated for training sessions, measured in minutes.

*Number of trials* (Ntrial): Total number of trials throughout the training.

*Training days* (Nday): Total count of training days included in the training program.

### 2.4 Trend analysis of variables of interest

To examine the popularity of variables included in our meta-analysis, we conducted a large-scale trend analysis of 34,772 studies on auditory category learning published between 2000 and 2024. We identified relevant articles through a comprehensive search that combined two groups of search terms: (1) auditory category-related terms (e.g., "auditory category") and (2) learning-related terms ("learning" or "training"). Abstracts from the identified papers were extracted and compiled into a single dataset. Utilizing Python 3.9 along with NumPy (version 2.1.0) and Pandas (version 2.2.3), we quantified the frequency of each variable by counting the annual occurrences of all associated keywords. Two variables, "sound type" and "number of trials," were excluded due to a lack of consistently used terms in the literature.

Below are the keywords used for defining each variable: talker variability ("recorder," or "talker," or "speaker"), context variability ("context," or "contextual," or "environment"), exemplar variability ("repeat," or "repetition"), instruction ("instruct," or "instruction"), presentation ("blocked," or "interleave"), feedback ("feedback," or "correct," or "wrong"), modality variability ("visual," or "cross-modal," or "gestures," or "video," or "modal," or "modality," or "gesture"), learning tasks ("active," or "passive"), duration ("duration," or "intensity"), and days ("days," or "sleep," or "consolidation", or "interval"). By tracking how often these terms appeared in the literature, we gained insight into their relative prominence in the field of auditory category learning over the years.

### 2.5 Effect size calculation

To provide a standardized method for measuring behavioral changes before and after training, and to facilitate comparisons across studies, we calculated effect sizes (Hedges' $g$) by extracting the means, standard deviations, and sample sizes from pre- and post-training tests in each experiment of the selected studies. When these statistics were unavailable, we used alternative measures, such as *F*-tests or *t*-tests, to estimate Hedges' *g*, which adjusts for small sample bias compared to Cohen's *d* (Hedges & Olkin, 1985). The calculation was performed using the "esc" package, which implements the formulas outlined in *Practical Meta-analysis* (Lipsey & Wilson, 2000). For example, when the t-statistic $t$ and the sample size $N$ are provided, the effect size is computed as follows:

$$ES = \frac{2t}{\sqrt{N}} \times (1 - \frac{3}{4\,(df) - 9})$$

Other formulas for calculating Hedges' $g$ from other statistics are provided in the Supplementary Materials (e.g., regression coefficients, chi-squared statistics).

After calculating the effect sizes from individual experiments, we adopted a random-effects model to estimate the overall effect size across studies. This model was chosen over a fixed-effects model because a heterogeneity test revealed significant variation among the included studies. In meta-analysis, within-study variance reflects the variability in each study's effect size estimate, largely influenced by sample size and measurement error, whereas between-study variance captures the variability in effect sizes resulting from study design. Accordingly, studies with larger sample sizes often yield smaller within-study variance due to more precise estimates, while smaller studies tend to show greater variance. We assessed heterogeneity using the $I^2$ statistic (Higgins et al., 2003), where higher $I^2$ values indicate a higher degree of between-study heterogeneity. All analyses were conducted in R using the "esc" package (Ben-Shachar et al., 2020, version 0.5.1),

"Metafor" package (Viechtbauer, 2010, version 4.4.0), and the "Meta" package (Balduzzi et al., 2019, version 7.0.0).

## 2.6 Factor validation and general linear modeling (GLM)

Before GLM and factor analysis, we assessed correlations among the variables of interest to reveal their interdependence. For each pair of variables, the choice of correlation measure depended on the type of variable. Associations between two continuous variables were quantified using Spearman's rank correlation coefficient (ρ; Spearman, 1904), which captures monotonic relationships without assuming linearity. Associations between two categorical variables were evaluated using Cramér's V, derived from $\chi^2$ contingency tests (Cramér, 1999). For numeric-categorical pairs, we estimated effect size using Eta squared ($\eta^2$) obtained from one-way ANOVA models (Olejnik & Algina, 2003). To allow for direct comparison across variable types, all association measures were converted to a standard metric, Hedges' *g* (Hedges, 1981), which provides a standardized effect size adjusted for small sample bias. The resulting values were compiled into symmetric association matrices alongside corresponding *p*-values. To account for multiple comparisons, we applied the Benjamini-Hochberg false discovery rate (FDR) correction to the upper triangle of the *p*-value matrix and symmetrized the results (Benjamini & Hochberg, 1995). Only associations with FDR-adjusted $p < 0.05$ were retained as significant. For interpretability, non-significant cells were masked in the effect size matrix prior to visualization. Significant associations were then displayed in a heatmap, with color intensity indicating the magnitude of Hedges' *g*.

Next, we used the General Linear Model (GLM) approach to assess the contribution of each variable to the effect size. Univariate GLMs were run independently for each variable. Each model was specified with effect size as the dependent variable and a single training variable as the predictor, using the base glm() function in R (version 4.4.1). This process was repeated across all 12 training variables to generate a comparative profile of how each factor individually contributes to the learning effect sizes.

To assess whether the imbalance in the number of speech versus non-speech studies affects the comparison of their effect sizes, we conducted a subsampling analysis. First, we performed separate one-sample *t*-tests for speech and non-speech studies to determine if their effect sizes were significantly greater than zero. Next, we used a Welch two-sample *t*-test to compare their effect sizes. To further investigate whether the unequal sample sizes (speech = 71, nonspeech = 33) influenced the between-group comparison, we performed a Monte Carlo subsampling analysis. Specifically, we randomly sampled 33 speech studies to match the non-speech sample size and then re-conducted the between-group *t*-test. This process was repeated 5,000 times, and we summarized the distribution of mean differences and *p*-values across all iterations.

We then conducted a Confirmatory Factor Analysis to validate the proposed structure, in which 12 variables of interest were mapped onto three latent constructs. The CFA was performed using the lavaan package (version 0.6.17), employing maximum likelihood estimation as the default method. We evaluated model fit by examining the Tucker-Lewis index (TLI), comparative fit index (CFI), standardized root mean square residual (SRMR), and root mean square error of approximation (RMSEA). We used the cutoffs recommended by Hu and Bentler (1999), which require TLI and CFI values greater than 0.95, SRMR values less than 0.08, and RMSEA values less than 0.06 to assess the model fit.

Finally, after confirming the underlying latent factors and their relationships with the observed variables, we constructed three separate GLMs to evaluate the collective contribution of the variability, engagement, and intensity-related variables to effect sizes. For each GLM, we included the observed variables associated with each latent factor as predictors. We extracted the significance of each model's overall fit and $R^2$ to measure the explanatory power of each latent factor in predicting learning effect sizes.

## 2.7 Activation likelihood estimation

While the behavioral meta-analysis highlighted the impact of various factors and variables on learning performance, the neuroimaging meta-analysis explores how these outcomes might be reflected in changes in brain activation. To achieve this, we identified and extracted activation coordinates from contrasts of fMRI studies that reported increases or decreases in brain activation following training. We conducted an ALE analysis to estimate the effect size of brain activation changes based on reported contrast coordinates using GingerALE (version 3.0.2; http://www.brainmap.org/ale/; Eickhoff et al., 2009).

Two types of contrasts were included: firstly, increased activation contrasts. These were typically derived from comparisons of post-training > pre-training (e.g., Deng et al., 2011; Zatorre et al., 2012), reflecting regions with increased activation after training. Additionally, contrasts capturing condition-by-session interactions (e.g., greater activation after training for the trained condition compared to the untrained one) were also incorporated, as they showed how brain activity changes over training blocks. We also included cross-sectional comparisons of learner versus non-learner groups, where brain regions show significant increases for the learner group compared to the non-learners after training (Yoo et al., 2007). Secondly, we included contrasts of decreased activation, operationalized as pre-training > post-training contrasts (De Souza et al., 2013; Newman-Norlund et al., 2006). They represented brain areas with reduced activations after training.

All reported coordinates were converted to MNI space using GingerALE's built-in transformation algorithm (icbm2tal). ALE is a kernel-based meta-analytic method that models reported activation foci as probability distributions, evaluating whether their overlap across experiments exceeds what would be expected by chance (Eickhoff et al., 2009; Turkeltaub et al., 2002). We conducted two independent ALE analyses to assess the brain activations associated with training, focusing on both increased and decreased activations following training. For the increased activation analysis, a total of 41 contrasts were used to generate coordinates that represented increased activation post-training. An ALE map was created by evaluating the activation likelihood for each voxel. In a separate analysis of decreased activation, 16 contrasts were included to identify regions with consistently reduced engagement after training, resulting in a corresponding ALE map. Next, each ALE map was compared against a null distribution obtained through random permutations to determine whether the observed clustering of activation foci was statistically significant beyond chance levels (Eickhoff et al., 2012). To control for multiple comparisons, we applied a cluster-level Family-Wise Error correction (FWE) at $p < 0.01$, with 1000 permutations to ensure robustness.

### 2.8 Sensitivity analysis and publication bias

A sensitivity analysis was conducted to assess the robustness of the overall effect size and to examine how the meta-analysis results change when a specific study was excluded iteratively. This leave-one-study-out approach assessed whether any single study disproportionately influenced the meta-analytic estimate. Substantial changes in the overall effect size upon exclusion would indicate potential instability in the findings.

To evaluate publication bias, we created a funnel plot to visualize the relationship between individual study effect sizes and their standard errors. Without publication bias, this plot should appear as a symmetrical inverted funnel: smaller studies (with larger standard errors) scatter widely at the bottom, while larger studies converge towards the true effect size at the top. After visual inspection, we statistically estimated the asymmetry using Egger's Test, which employs a linear regression model that regresses effect sizes on standard errors. A slope significantly different from zero suggests a statistically significant bias. We also applied the Trim and Fill Method (Shi & Lin, 2019) to adjust for publication bias by imputing potentially missing studies. This method trims the most extreme effect sizes on one side of the funnel plot and fills in missing studies by estimating their effect sizes based on the remaining data, ultimately recalculating the overall effect size with both observed and imputed studies.

To assess the potential impact of publication bias in the neural data, we conducted a Fail-safe N analysis (Orwin, 1983), which estimates the number of null studies required to alter the significant findings to non-significance. To achieve this, null experiments with randomly sampled peak activations, which approximate the sample size and number of foci of included papers, were generated with a publicly available R script (https://github.com/NeuroStat/GenerateNull). Then, the simulated foci were added iteratively to the original experiments to estimate the maximum number of unpublished null studies required to make each cluster non-significant (Acar et al., 2018). The Fail-safe N analysis was performed for the pre-post and post-pre contrasts separately. We considered a result robust if this number exceeded established thresholds: a lower boundary set at 30% of the included studies (based on previous estimates suggesting roughly 30 unpublished null studies per 100 studies (Samartsidis et al., 2020) and an upper boundary set at five times the number of included studies (Enge et al., 2021).

### 2.9 Lateralization analysis

Given the distinct roles of the left and right auditory pathways, lateralization analysis can provide insights into whether one hemisphere is more actively involved in auditory category learning. The lateralization index also closely relates to our behavioral variable coding of sound type (Speech vs. Non-speech). The left hemisphere is well-established as the primary locus for speech processing and categorical perception (Poeppel et al., 2004; Studdert-Kennedy & Shankweiler, 1970), while the right hemisphere is more engaged in processing spectral and prosodic information (Friederici & Alter, 2004; Tervaniemi & Hugdahl, 2003). By examining lateralization, we can determine if one hemisphere has a greater role in learning auditory categories across speech and non-speech domains, highlighting neural specialization during the learning process.

We used the Standardized Lateralization Index (SLI) (Dietz et al., 2016) to quantify the hemispheric asymmetry of brain clusters involved in auditory category learning. The SLI is computed as follows:

$$SLI = \frac{\text{Left Active Volumes} - \text{Right Active Volumes}}{\text{Left Active Volumes} + \text{Right Active Volumes}}$$

This formula produces an SLI value ranging from -1 to 1. Negative values indicate greater right-hemisphere activation, while positive values reflect greater left-hemisphere activation. Following Szaflarski et al. (2005), an SLI between -0.1 and 0.1 suggests bilateral activation, SLI > 0.1 indicates left lateralization, and SLI < -0.1 indicates right lateralization. We performed the lateralization analyses for the two types of contrasts separately.

## 2.10 Functional decoding

We employed the functional decoding approach (Poldrack, 2011) to infer the cognitive processes associated with the areas identified by ALE. Functional decoding uses annotated neuroimaging databases to link regional activations with study-level metadata, clarifying the functional roles (e.g., learning, memory, or auditory processing) of the implicated brain areas. Specifically, this process begins with a thresholded ALE image derived from prior meta-analysis, which is then converted into a binary mask. Using the Automated Anatomical Labeling (AAL) atlas, seven predefined regions of interest (ROIs) were selected, including bilateral regions such as Heschl's Gyrus (HG), Superior Temporal Gyrus (STG), Inferior Frontal Gyrus (IFG), Precentral Gyrus (PreCG), Insula, Cuneus, and Precuneus. These ROIs intersected with the ALE mask to isolate ROI-specific activation clusters. These ROI-constrained activation clusters were then input into NiMARE's ROIAssociationDecoder, which computed the association between the masked activation and the reference database, NeuroSynth (Yarkoni et al., 2011). The decoder produces a ranked list of cognitive terms (e.g., "auditory", "language", "memory") that are statistically associated with each cluster. All functional decoding procedures were performed using the NiMARE library (version 0.2.0; Salo et al., 2023).

## 3. RESULTS

### 3.1 Topic trend and variable distribution

The topic trend analysis showed a significant upward trend in research on training variables from 2000 to 2024 (Figure 2a). Most variables have shown increased research attention over time, with a notable acceleration between 2010 and 2015. The total frequency (scaled down by 10) confirms the overall expanding research interest in this field. Among the terms examined, those concerning training variability were the most frequently studied. Context variability has experienced the most significant increase, surpassing 1000 mentions by 2024, with particularly rapid growth after 2017. Engagement-related terms, such as learning task and feedback, also showed moderate and steady growth in research frequency over the years. Intensity-related terms had comparatively lower frequencies.

The distribution and association of training variables (see Figures 2b and 2c) reveal distinct patterns and connections between different experimental factors. Four relationships among training variables in behavioral studies were worth highlighting. First, a strong pairing exists between context variability and talker variability, illustrated by the Sankey diagram with thick flow lines showing the connection (Figure 2b). The correlational analysis confirmed a significant positive correlation (Hedge's $g = 0.395$, $p = 0.001$), indicating that using multiple talkers is often associated with manipulating variability in phonetic contexts (Figure 2d, upper left). Second, context variability correlates with instruction type (Hedge's $g = 1.829$, $p = 0.001$), where higher context variability aligns with more comprehensive instructions (Figure 2d, upper right), suggesting that diverse learning contexts may co-occur with explicit instructional support. Third, feedback implementation varies across modality variability levels, with full feedback more commonly linked to multimodal presentations (Hedge's $g = 2.154$, $p = 7.650\times10^{-4}$; Figure 2d, lower left). Finally, a notable interaction exists between modality variability and learning task type (Hedge's $g = 2.803$, $p = 5.374\times10^{-7}$, Figure 2d, upper right). Multimodality scenario typically involves active learning, while unimodal often corresponds with passive learning paradigms, as indicated by a significant correlation coefficient that shows how researchers align task engagement with presentation diversity.

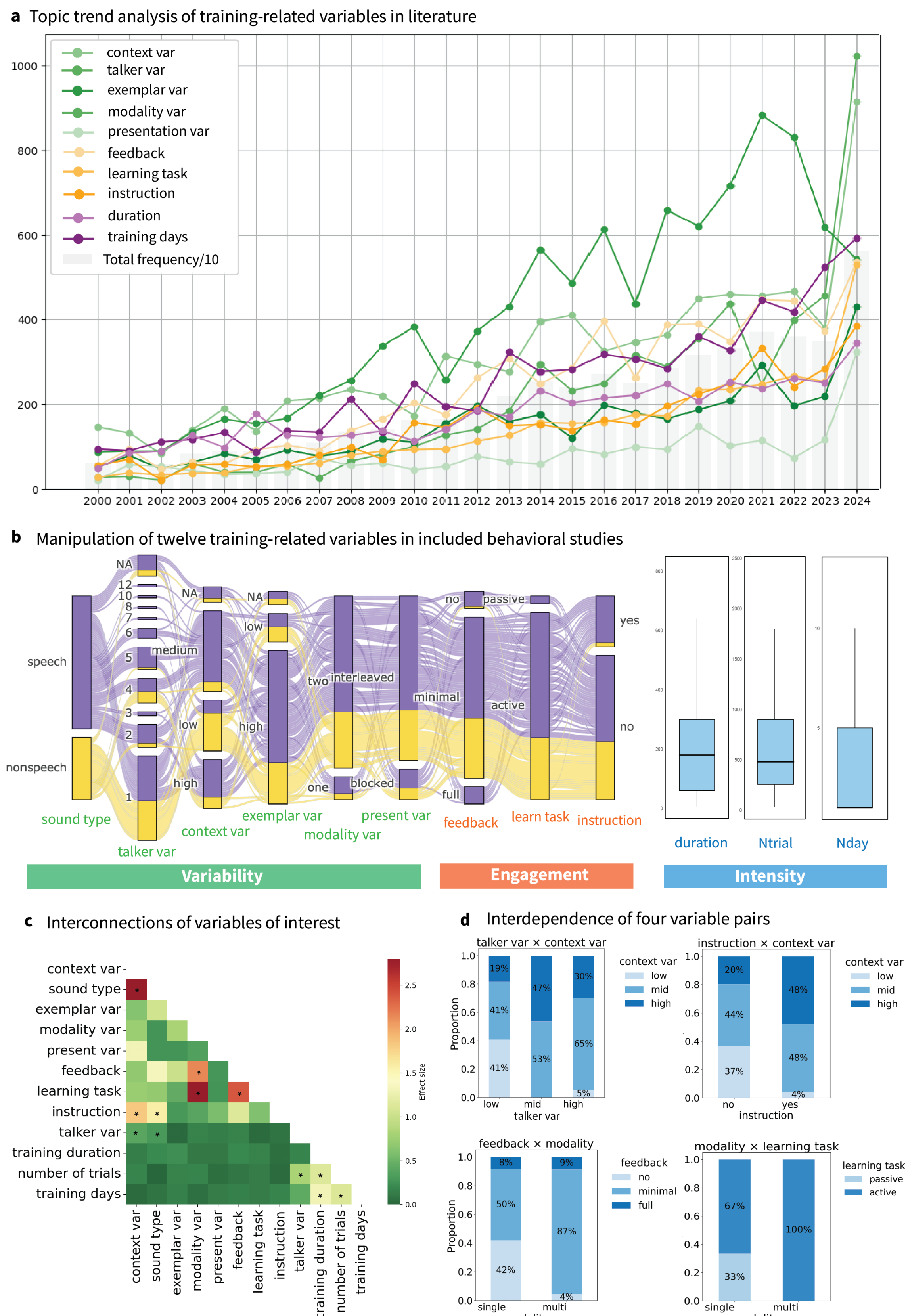
a Topic trend analysis of training-related variables in literature
context var
talker var
exemplar var
modality var
presentation var
feedback
learning task
instruction
duration
training days
Total frequency/10
0
200
400
600
800
1000
2000 2001 2002 2003 2004 2005 2006 2007 2008 2009 2010 2011 2012 2013 2014 2015 2016 2017 2018 2019 2020 2021 2022 2023 2024
b Manipulation of twelve training-related variables in included behavioral studies
speech
nonspeech
NA
12
10
8
7
6
5
4
3
2
1
medium
low
high
two
one
interleaved
blocked
minimal
no
full
passive
active
yes
sound type
talker var
context var
exemplar var
modality var
present var
feedback
learn task
instruction
duration
Ntrial
Nday
Variability
Engagement
Intensity
c Interconnections of variables of interest
context var
sound type
exemplar var
modality var
present var
feedback
learning task
instruction
talker var
training duration
number of trials
training days
Effect size
d Interdependence of four variable pairs
talker var × context var
instruction × context var
feedback × modality
modality × learning task
Proportion
context var
low
mid
high
feedback
no
minimal
full
learning task
passive
active
talker var
instruction
modality var
single
multi
19%
41%
41%
47%
53%
30%
65%
5%
20%
44%
37%
48%
48%
4%
8%
50%
42%
9%
87%
4%
67%
33%
100%

**Figure 2.** Study trends and connections between training variables. (**a**) The research trends for the 12 variables were grouped into three factors. Variables related to engagement are shown in orange, those associated with variability are in green, and those connected to intensity are in blue. The grey bar represents the total annual frequency of all topics, scaled down by a factor of ten for visualization. (**b**) A Sankey plot and three boxplots illustrate the distributions of variables across the included studies, highlighting their interconnections. Each column in the diagram corresponds to a specific variable, with colored flows determined by the first column, which represents the sound type. The thickness of these flows reflects the relative proportion of studies that follow each path. Each flow connects these columns and illustrates how the levels of one variable relate to those of others. The box plots on the right quantify the distribution of the three intensity variables, which strongly correlate with one another. The three colored boxes beneath the figure (variability, engagement, and intensity) represent the grouping factor of the twelve observed variables. Values shown as "NA" indicate that the corresponding information was not reported in the original study. The visualizations were created using RcolorBrewer (version 1.1.3), easyalluvial (version 0.3.2), parcats (version 0.0.5), and ggplot2 (version 3.5.1) in R (version 4.4.1). (**c**) Heatmap showing pairwise associations among training variables. Associations were measured using Spearman's ρ (numeric-numeric), Cramér's V (categorical-categorical), or $Eta^2$ (numeric-categorical), and converted to Hedges'*g* for comparisons across variable pairs. False discovery rate correction (FDR, Benjamini-Hochberg) was applied to all p-values, and only significant associations (FDR-adjusted $p < 0.05$) are marked with asterisks. The exact *p*-values are provided in Supplementary Table S6. (**d**) Stacked bar plots showing the distribution and interdependence of four categorical variables. Talker variability × Context variability: Proportion of experiments grouped by talker variability bins (low, mid, high) and stacked by context variability (low, mid, high); Instruction × Context variability: Proportion of experiments for instruction conditions (no vs. yes), stacked by context variability (low, mid, high); Feedback × Modality: Proportion of experiments for feedback conditions (no, minimal, full), stacked by modality (single vs. multi); Modality × Learning task: Proportion of experiments for modality conditions (single vs. multi), stacked by learning task (passive vs. active).

### 3.2 Overall effect size of auditory category learning

The behavioral meta-analysis comprised 104 experiments. The overall estimated effect size was Hedges' $g = 1.45$ ($SE = 0.09$, 95% $CI$ [1.28, 1.62]), indicating a significant difference from chance ($z = 16.59$, $p < 0.0001$). Effect sizes are generally categorized as small (0.2), medium (0.5), and large (above 0.8), according to Hedges & Olkin (1985). This overall effect size suggests a substantial improvement in behavior following training. We also found a large variation in behavioral effect sizes across studies (see Figure S1 for effect sizes of individual studies). We performed a heterogeneity test to assess whether the observed differences in effect sizes across studies were due to sampling error or variations resulting from between-study differences (e.g., experimental design). This test yielded significant results ($p < 0.0001$) with an $I^2$ value of 91.39%, which exceeds the high heterogeneity threshold of 75% set by Higgins (2003). This suggests that most variations across studies are attributable to between-study differences rather than sampling errors.

### 3.3 Distinct variable contribution to auditory category learning

We further examined the contribution of each variable to the learning effect sizes. The effect of each variable on effect sizes is visualized in multiple box plots in Figure 3 (see Table S4 for detailed statistical results).

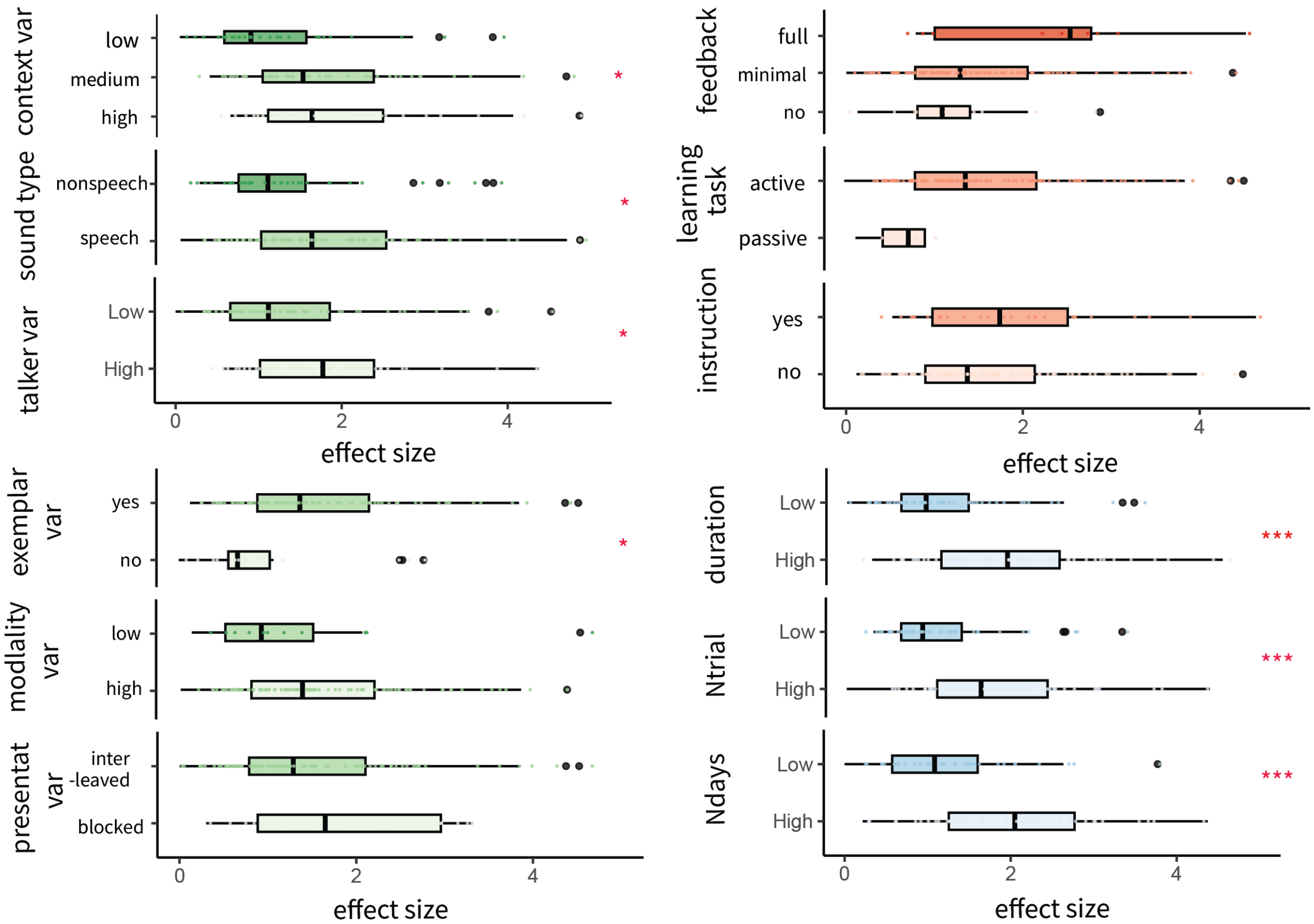

**Figure 3**. The distribution of effect sizes at different levels of each variable of interest. Three distinct colors denote the three hidden factors: green for variability, orange for engagement, and blue for intensity. The three training intensity variables are median-split into two levels for visualization. Statistically significant differences between levels were annotated with asterisks in the plots, with * indicating $p < 0.05$ and *** indicating $p < 0.001$.

Using GLM, we identified several variables that significantly predicted effect sizes. Both the number of trials and training duration (jointly reflecting training intensity) were significant, with beta coefficients of 0.267 ($p = 0.011$) and 0.247 ($p = 0.035$), respectively. Regarding variability-related variables, talker variability (beta = 0.108, $p = 0.018$), context variability (beta = 0.36, $p = 0.007$) and sound type (beta = 0.46, $p = 0.026$) showed robust positive effects, suggesting that greater exposure to multiple talkers and contexts supports better learning outcomes. Exemplar variability yielded a negative coefficient of beta = -0.559 ($p = 0.043$), indicating that repeated exposure to the same training item can promote learning. Among engagement-related variables, feedback emerged as a significant predictor of improvement (beta = 0.511, $p = 0.028$). Other variables (i.e., modality variability and learning task) demonstrated positive but non-significant effects on the learning effect size.

A confirmatory factor analysis was subsequently used to validate the grouping of the 12 variables into three latent factors. Each factor model demonstrated a good overall fit. The variability latent factor model fit well (CFI = 0.941, TLI = 0.902, RMSEA = 0.070, SRMR = 0.149), where both talker variability ($\lambda = 2.77$, $p = 0.018$) and sound type ($\lambda = 2.06$, $p = 0.012$) loaded significantly on the variability factor, with context variability constrained as a reference indicator. The intensity model was robust (CFI = 1.000, TLI = 1.000, RMSEA = 0.000, SRMR = 0.000), with significant loadings for the number of trials and training days (ranging from 1.12 to 1.27). Although the engagement model also showed good fits overall (CFI = 1.000, TLI = 1.000, RMSEA = 0.000, SRMR = 0.000), no variable loadings reached significance.

To examine the collective predictive power of the variables from each latent factor on effect sizes, we constructed three separate GLMs for the intensity, variability, and engagement variables (Figure 4b). Results revealed that the intensity variables jointly explained the largest proportion of variance in effect sizes ($R^2 = 0.37$, $p = 1.29 \times 10^{-6}$; Figure 4b, right panel), followed by the variability factor, which contributed a moderate yet significant amount of explained variance ($R^2 = 0.19$, $p = 0.002$; Figure 4b, left panel). The engagement model was marginally significant and accounted for a small portion of variance ($R^2 = 0.06$, $p = 0.07$; Figure 4b, middle panel). Overall, these findings highlight the prominent roles of intensity and variability in driving auditory category learning success, whereas engagement appears to play a weaker, though potentially still influential, role in modulating auditory category learning.

We found that the behavioral dataset included more speech studies than non-speech studies. To examine whether this might bias results, we conducted further analysis. Both speech and non-speech studies showed effect sizes significantly greater than zero. Speech studies had a mean effect size of 1.67, $t = 13.85$, $p < 2.2 \times 10^{-16}$, and non-speech studies had a mean effect size of 1.21, $t = 8.11$, $p = 2.91 \times 10^{-9}$. A direct comparison between the two groups indicated a significant difference in effect sizes, Welch's $t = -2.39$, $p = 0.019$. We further used subsampling analysis to balance the sample size and re-conducted the Welch's *t*-test 5000 times. We found 48.5% of iterations yielded a significant group effect with the threshold of $p = 0.05$. The median mean difference was 0.46, and the median *p*-value was 0.052. Taken together, these results demonstrate that both speech and non-speech auditory category learning studies produce strong above-chance effects, while Sound type has a marginally significant impact on the learning effect size when the number of included studies is balanced.

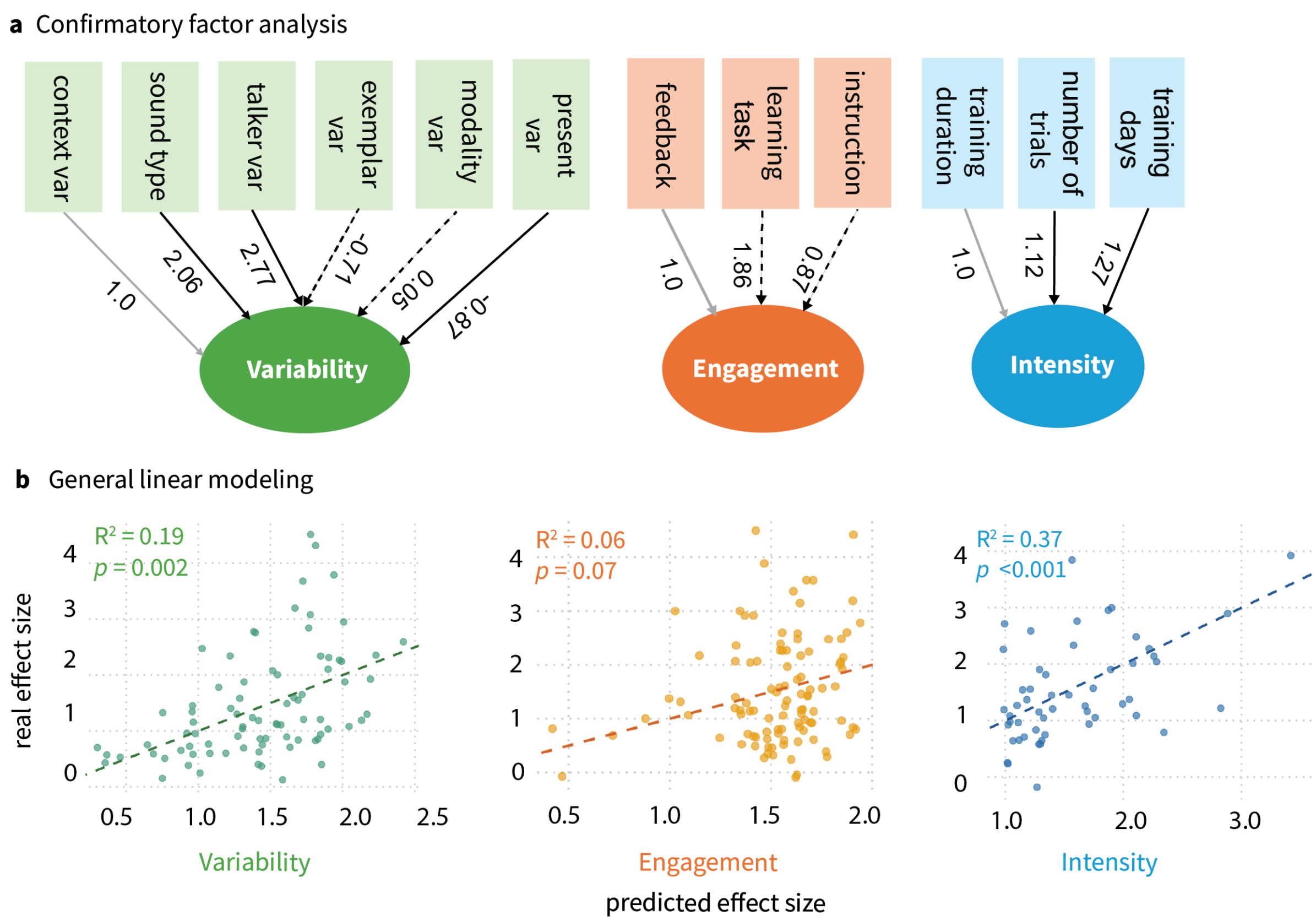

**Figure 4.** Validation of the three latent factors and their contributions to learning effect sizes. (**a**) Results of confirmatory factor analysis for the three latent factors. Factor loadings for the 12 training variables are shown along the paths, with reference indicators fixed at 1.0 (grey lines). Solid black lines indicate significant paths. (**b**) Scatter plots show predicted versus observed effect sizes, with regression lines, $R^2$ values, and $p$-values.

### 3.4 Brain activation changes following auditory category learning

To investigate brain activation changes linked to auditory category learning, we conducted ALE analyses on coordinates reported for increased (post- > pre-training) and decreased (pre- > post-training) activations separately (see Figure 5a for all activation coordinates). Our findings revealed that brain activation changes in a left-lateralized auditory frontotemporal-parietal network supported auditory category learning.

Two main clusters showed significantly increased activations (i.e., post > pre; see Figure 5b). The first was left-lateralized, centered in the Transverse Temporal Gyrus (Heschl's Gyrus, BA 41; peak coordinates: [-52, -18, 10]), with extended regions to the Insula (BA 13), Precentral Gyrus (BA 6), and Inferior Frontal Gyrus (BA 9, 44) (Figure 5b, left panel). A second, right-lateralized cluster emerged in the Superior Temporal Gyrus (BA 22; peak coordinates: [52, -4, 0]) and Precentral Gyrus (BA 6; peak coordinates: [56, 0, 10]) (Figure 5b, right panel). Additionally, we observed that a cluster from the Cuneus (BA 7; peak coordinates: [-6, -72, 38]) to the Precuneus (BA 31; peak coordinates: [-8, -68, 26]) showed significant decreases in activation (i.e., pre > post) following training (See Supplementary Table S2 for details). A lateralization analysis with SLI = 0.177 confirmed a left-hemisphere dominance for auditory category learning-related activation changes.

Functional decoding results illustrated in Figure 5c reveal distinct functional profiles for the identified brain clusters. The frontotemporal cluster (HG, STG, IFG, and Insula) shows strong associations with auditory-language-related processes, such as speech, sound, and music. The Cuneus and Precuneus clusters, which exhibit a pattern of decreased activation, were associated with memory retrieval and monitoring processes. These findings highlight the role of a left-lateralized auditory pathway in supporting auditory learning, while also demonstrating that bilateral temporal areas and other cognitive components contribute to auditory category learning.

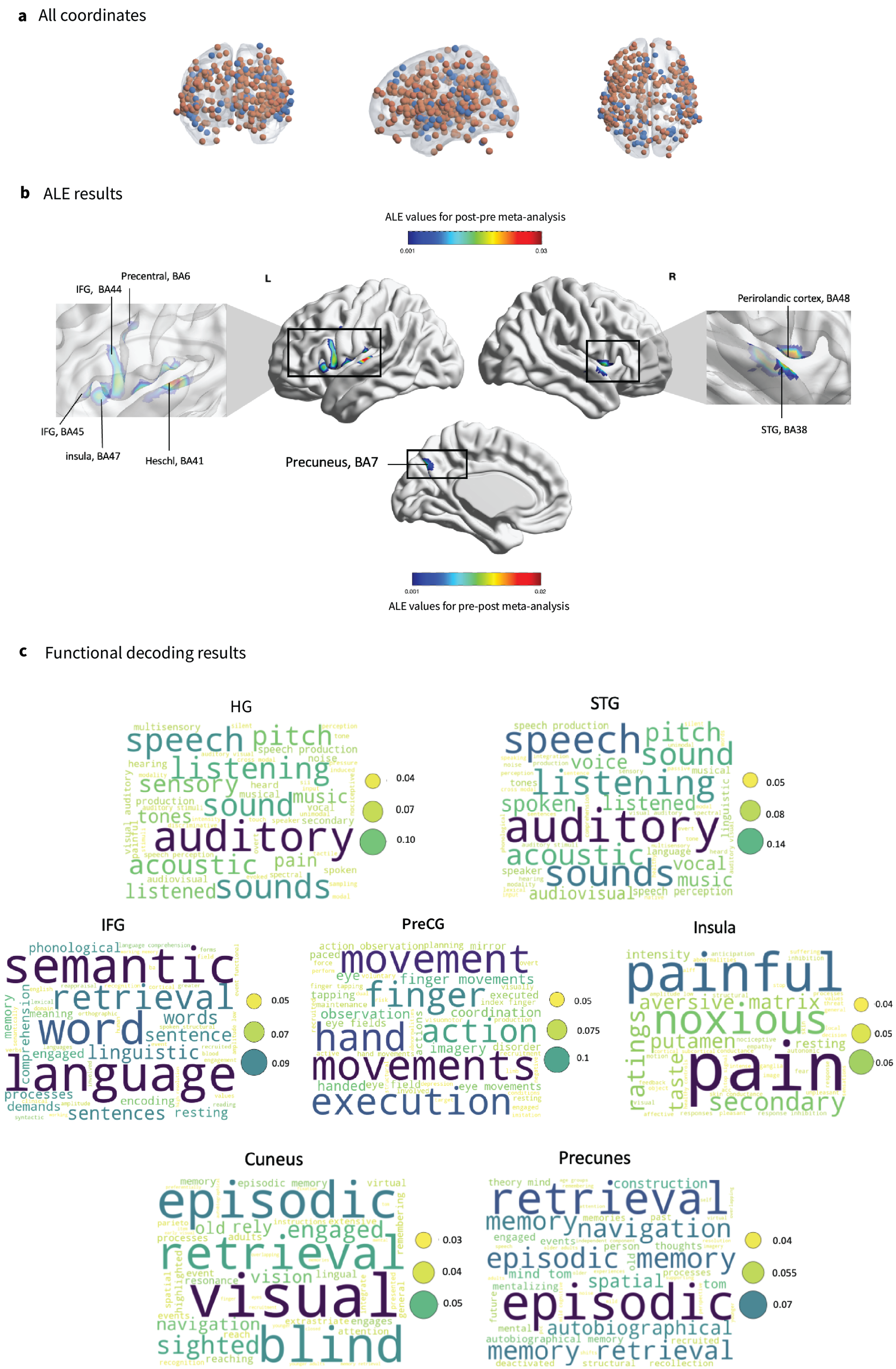
a All coordinates
b ALE results
ALE values for post-pre meta-analysis
0.001
0.03
L
R
Precentral, BA6
IFG, BA44
IFG, BA45
insula, BA47
Heschl, BA41
Perirolandic cortex, BA48
STG, BA38
Precuneus, BA7
0.001
0.02
ALE values for pre-post meta-analysis
c Functional decoding results
HG
speech pitch listening sensory sound music tones auditory acoustic pain listened sounds
0.04
0.07
0.10
STG
speech pitch sound voice listening spoken listened auditory acoustic vocal sounds music audiovisual
0.05
0.08
0.14
IFG
semantic retrieval word words sentence linguistic language sentences processes demands
0.05
0.07
0.09
PreCG
movement finger movements finger hand action movements execution coordination observation
0.05
0.075
0.1
Insula
painful aversive matrix noxious putamen pain secondary ratings taste
0.04
0.05
0.06
Cuneus
episodic retrieval visual blind sighted navigation vision engaged
0.03
0.04
0.05
Precunes
retrieval memory navigation episodic memory episodic autobiographical memory retrieval spatial
0.04
0.055
0.07

**Figure 5**. Brain activation changes following auditory category learning. (**a**) A total of 372 foci extracted from 57 contrasts and 1242 subjects are projected onto a 3D brain model. Red nodes represent the 296 coordinates with increased activation after training, while blue nodes correspond to the 76 coordinates with decreased activation after training. (**b**) ALE results were visualized using BrainNet Viewer (Xia et al., 2013). The upper panel highlights increased activation in HG, STG, IFG, PreCG, and Insula, while the lower panel shows decreased activity in medial parietal areas (Cuneus and Precuneus). (**c**) Functional decoding results show word clouds associated with each of the seven regions identified in the ALE analysis. Each word in the cloud corresponds to a term related to the region in the NeuroSynth Dataset, with the word's color indicating its strength of association.

### 3.5 Robustness analyses and publication bias estimation for behavioral and neural meta-analyses

To ensure the reliability of our findings, we conducted sensitivity analyses and assessments of publication bias for both behavioral and neural data. Due to significant heterogeneity in behavioral effect sizes ($I^2$ = 91.39%, $p < 0.0001$), we performed a leave-one-study-out sensitivity analysis. Removing individual studies yielded effect sizes ranging from 1.28 to 1.62, all of which were statistically significant with 95% confidence intervals that excluded zero, indicating that no single study skewed the overall effect size. A funnel plot comparing standard errors with Hedges' *g* showed marked asymmetry, with smaller studies clustered on the right side (see Figure S2 in Supplementary Materials). Egger's test confirmed this asymmetry ($t = 6.01$, $p < 0.001$), suggesting publication bias. The trim-and-fill method imputed 31 potentially missing studies, resulting in a standardized mean difference of 0.96, indicating that the original estimate might be inflated by publication bias.

Similar robustness analyses were conducted for the ALE meta-analysis. In the post-pre cohort, reproducibility rates were consistently high, showing varying degrees across regions. The left HG demonstrated robust reliability in 97.5% of cases, the Insula in 92.7%, the IFG in 95.1%, and the PreCG achieved a 100% reproducibility rate, while the right STG had a rate of 85.4%. For the pre-post cohort, a leave-one-study-out analysis revealed that clusters in the Cuneus and Precuneus were reproducible in 75% of iterations (12 out of 16). These results emphasized strong left-dominant neural changes and relatively reduced stability for the Cuneus and Precuneus.

## 4. DISCUSSION

Our meta-analysis offers new insights into how latent factors of training variability, intensity, and engagement influence the success of auditory category learning, providing a more comprehensive and data-driven perspective than earlier studies that focused on individual variables. Our findings align with established principles of experience-dependent plasticity, where variability and intensity are recognized as key factors in general learning and neural plasticity (Kleim & Jones, 2008). Our meta-analysis makes a new contribution by situating these factors within the domain of auditory category learning and integrating them into a three-dimensional framework of variability, intensity, and engagement. While previous research has examined these factors separately, no prior work has systematically examined how they jointly contribute to behavioral outcomes and neural changes across a large number of studies. By synthesizing results from 111 behavioral experiments, we showed that training intensity and variability significantly contribute to predicting learning success and have a more prominent role in auditory category learning than engagement. Additionally, our ALE meta-analysis of neuroimaging data indicates that successful learning is linked to increased activation in a left-lateralized frontotemporal network and reduced activation in memory-related regions (i.e., cuneus and precuneus). This may suggest a transition from relying on memory to more efficient abstract categorization and generalization processes that depend on cortical auditory-language pathways following training. These findings improve our understanding of auditory category learning success by analyzing multiple training variables, integrating behavioral and neural findings, introducing a new framework for categorizing training variables, and providing evidence-based insights to optimize training protocols.

### 4.1 Distinct contribution of training variables to behavioral learning

A key objective of this meta-analysis was to evaluate the extent to which these training variables, both individually and collectively as a latent factor, predict the effect sizes of auditory category learning. Addressing this multifactorial nature of auditory category learning is challenging in individual experiments but becomes more tractable through integrative meta-modeling across numerous studies.  By combining confirmatory factor analysis and general linear modeling, we evaluated the influence of 12 candidate variables, categorized into three latent factors, on behavioral learning effects.

Several variables (i.e., talker variability, context variability, feedback, exemplar variability, training duration, and number of trials) were positively correlated with increased effect sizes. These findings resolve some seemingly contradictory conclusions in earlier studies. For instance, high variability has sometimes been found to both facilitate and hinder learning, likely depending on a learner's proficiency and stage of learning (Perrachione et al., 2011) or the complexity of the materials (Sadakata & McQueen, 2013). Our broad-scale data indicate that variability is generally beneficial under many conditions. With respect to feedback, prior research has shown that its benefits are not uniform. For example, McCandliss et al. (2002) found that the usefulness of feedback may vary depending on the perceptual space to be reshaped by newly learned categories, specifically in their testing case in phoneme training (categorizing /l/ and /r/) for Japanese speakers. Our meta-analysis suggests that providing feedback generally facilitates learning, but different types of feedback may depend on the learning context and category structure.

Consistent with theoretical and empirical work highlighting the role of exposure (e.g., Estes & Burke, 1953; Kartushina & Martin, 2019) and variability (e.g., Brekelmans et al., 2022) in auditory learning, our results show that intensity (e.g., training duration and number of trials) and variability (e.g., training with multiple talkers and contexts) exert the strongest overall impact on behavioral performance. The significance of variability as a mechanism for generalization has been widely tested; however, the substantial effect of intensity highlights the practical necessity of ensuring adequate training "dosage." This emphasis on exposure, concerning both duration and frequency of training, introduces a possible intensity-by-variability interaction mechanism into ongoing discussions: even rich and diverse input may not provide optimal benefits unless learners also have sufficient time and repeated practice to reinforce their auditory category knowledge (rules or boundaries). In addition to emphasizing variability in practice, the amount of exposure (i.e., duration and frequency of training) critically affects outcomes. By illustrating that training intensity can be as influential as variability, these findings encourage educators and researchers to balance both factors, input diversity and adequate practice time, when designing training protocols and educational interventions.

The Engagement factor showed weaker loadings than Variability and Intensity, likely due to several reasons. First, feedback was the only engagement-related variable significantly predicting learning success, unlike learning task and instruction, which did not reach significance, suggesting feedback has a more direct influence on performance. Second, Engagement is a more abstract and diverse construct than Variability and Intensity, which can be relatively easily measured by metrics like the number of talkers or training duration. Engagement includes broad elements such as task type, instruction explicitness, and feedback, which vary widely and are inconsistently implemented across studies, making it less cohesive as a single construct. Third, Figure 2B shows an uneven distribution of Engagement variables across studies, further weakening factor loadings. Overall, the weaker Engagement factor reflects its conceptual diversity and uneven study representation. Future work should develop new engagement measures and ensure balanced study designs.

Significant predictors of behavioral effect size were identified among variables related to training variability, exemplar variability, feedback, talker variability, and context variability. Among these, talker and context variability demonstrated the strongest positive effects compared to other forms of variability (e.g., modality variability, presentation sequence). This may be because talker and context variability provide highly relevant acoustic cues that enhance generalization without overwhelming learners with irrelevant changes. Importantly, research on category learning shows that such variability can speed up the discovery of category-relevant dimensions by emphasizing consistent features across different talkers or contexts and reducing dependence on unique cues (Gómez, 2002; Lively et al., 1993; Logan et al., 1991). Variability thus not only helps expand generalization but also guides learners' attention to the most diagnostic features of the inputs. In line with this speculation, research on voice recognition (Goh, 2005) suggests that voice-specific details are encoded into long-term memory, highlighting how variability in speaker identity enriches learners' perceptual representations. This deeper encoding promotes more robust and generalized category knowledge, underscoring the importance of strategically incorporating different types of variability into training regimens.

### 4.2 Auditory frontotemporal-parietal network subserves auditory category learning success

Through a meta-analysis of brain coordinates from 29 fMRI experiments, we identified a left-lateralized frontotemporal network that shows a significant increase in activity after auditory category learning. This network particularly activates in auditory-language and motor-related regions, including the left inferior frontal gyrus (IFG), left Heschl's gyrus (HG), right superior temporal gyrus (STG), insula, and precentral gyrus (PreCG). In contrast, we noted a decrease in activity in memory-related areas, especially the cuneus and precuneus.

The left HG likely subserves fine-grained spectro-temporal analysis of phonemes, whereas the right STG is more closely linked to processing tonal or prosodic features of auditory stimuli (Hickok & Poeppel, 2007). This hemispheric distinction receives further support from other meta-analytic findings. Liang & Du (2018) reported that lexical-tone perception and prosody tend to elicit stronger engagement in the right auditory cortex, whereas phoneme perception is more left-lateralized. Alongside these perception-oriented areas, the inferior frontal areas were also activated, echoing work demonstrating the crucial role of auditory-motor coupling in learning novel sound categories (López-Barroso & de Diego-Balaguer, 2017). The PreCG, meanwhile, is often linked to articulatory gestures (Pulvermüller et al., 2006), and thus auditory-motor mapping may facilitate bridging perception with motor representations in auditory category learning.

An additional intriguing finding was a convergence of decreased activity in the cuneus and precuneus. According to functional decoding results and prior literature, the precuneus is part of episodic or autobiographical memory networks, which are often more active during retrieval than encoding (Cavanna & Trimble, 2006; Daselaar et al., 2009). This result diverges from some earlier perceptual learning studies in which the sensory cortex showed primary downscaling of activity (Censor et al., 2006; Watanabe & Sasaki, 2015; Yotsumoto et al., 2008). One explanation may be the nature or stage of auditory category learning; in earlier phases, reliance on stored sensory exemplars can be high (Ohl et al., 2001), whereas later, speech-auditory and motor-related regions more automatically integrate these newly acquired acoustic distinctions. Thus, as learners become more skilled at differentiating categories, their cognitive demands shift from memory-based mechanisms to more automatic, skill-based processes (Kirschen et al., 2005; Lohse et al., 2014; Stein et al., 2009). The reduced precuneus and cuneus engagement may reflect a transition away from explicit recall and toward more efficient perception-motor mapping. However, the robustness of this cuneus-precuneus convergence is constrained by the relatively small number of contrasts (16 studies), which is just below the recommended cutoff proposed by Eickhoff et al. (2016). Therefore, the decreased activations in the two regions contributing to auditory category learning success should be approached with caution.

Compared to previous meta-analyses in auditory category perception (Liang & Du, 2018), our findings indicate that both category learning and category perception broadly recruit the STG and the PreCG. However, learning involves greater engagement and change in the IFG and insula, along with suppression in the cuneus and precuneus. Notably, our meta-analysis did not show significant recruitments in subcortical regions such as the striatum and cerebellum that Dual Learning System models emphasize (Krishnan et al., 2016). This discrepancy possibly arises from sample characteristics (e.g., tasks placing less emphasis on procedural or reward-based elements) or the dilution of effect sizes within broad meta-analytic approaches (Lim et al., 2019; Schönberg et al., 2007). Another possibility is that our neural measure is only focused on the end products of learning by using post- vs. pre-training neural activation changes, while the learning system proposed by the Dual Learning System models is more focused on the process of learning. Thus, future studies should pay more attention to linking the learning process with the outcome of learning.

The spatially localized networks identified in our fMRI analysis, particularly the left-lateralized frontotemporal network, could be further studied using EEG and MEG to track neural synchrony and oscillatory activity during the learning process. Previous EEG and MEG studies have revealed that neural oscillations in the theta (4-8 Hz) and gamma (30-100 Hz) frequency bands play crucial roles in auditory category learning. Theta oscillations are typically associated with higher-order integration, such as memory encoding and retrieval, while gamma oscillations are linked to early sensory processing and the integration of features across multiple acoustic dimensions (Kaiser & Lutzenberger, 2005; Kobayashi et al., 2024; Roux & Uhlhaas, 2014; Tang et al., 2016). Given that our findings suggest a shift away from reliance on memory retrieval toward automatic, skill-based processes involving sensory-motor integration, EEG studies could complement this observation by tracking how oscillations in the two frequency bands change and interact over time. Also, as we observed activation in both STG and PreCG for auditory category learning, EEG and MEG studies could build on these results by examining neural oscillations in these regions, to explore how sensory processing (e.g., in the STG) and motor mapping (e.g., in the PreCG) unfold across different stages of auditory category learning.

### 4.3 Limitations and future directions

Our current modeling mostly validates the proposed intensity-variability-engagement framework for auditory category learning; this result may stem from the relatively limited number of studies in our dataset. To address this, future investigations should include larger samples and employ more advanced analytical methodologies to thoroughly evaluate these three putative dimensions of auditory category learning. Especially, the limited sample of 29 neuroimaging studies warrants caution, given the relatively small number of contrasts (16 studies) for the decreased activations; it may not fully capture the depth of neural processes involved in auditory category learning or reveal subtle subgroup differences (e.g., implicit versus explicit paradigms). Additionally, since the current meta-analysis only collected data from healthy adult participants, our conclusions might not fully generalize to other age groups and could leave important developmental differences unexplored in how auditory categories are learned. For example, infants and young children may be more sensitive to statistical patterns, such as transitional probabilities between syllables and phonetic cues in speech, during early speech sound learning, which helps them quickly segment and categorize auditory input with great efficiency (e.g., Kuhl, 2004a; Saffran et al., 1996). This heightened sensitivity is thought to stem from greater neural plasticity in early childhood, a period when the brain's perceptual systems are highly adaptable and passively respond to surrounding linguistic patterns, often outperforming adults in implicit learning tasks (Kuhl et al., 2006; Werker & Hensch, 2015). Conversely, adults tend to depend more on conscious strategies and show less flexibility because of native-language biases, while older adults might experience further declines in auditory processing and plasticity due to age-related brain changes (e.g., Anderson et al., 2012; Wong et al., 2009). These findings collectively suggest that the mechanisms underlying auditory category learning, including statistical inference and perceptual tuning, may differ across various age groups. Future research should systematically explore different factors and paradigms with younger groups (e.g., infants and children) and older adults. Such studies could reveal how learning, plasticity, and real-world applications, such as improving language therapies for developmental issues or age-related cognitive decline, vary across the lifespan.

Another limitation is that our operationalization of talker and context variability was speech-centric. These variables captured phonetic context and talker variability, suitable for speech but not for non-speech studies. Despite similar forms of variability such as instrument variability, environmental sound changes, or device variability in artificial stimuli, which could be relevant.

However, since most non-speech studies used synthesized tones or non-speech signals with frequency modulation, and none involved musical instruments, measuring comparable variability across speech and non-speech studies was challenging. We detailed stimulus type in Supplementary Table S1 to help readers evaluate the variability in non-speech category studies. Future meta-analyses should adopt domain-general coding schemes, such as viewing "sound-source variability" as a higher-order construct to better capture variability across both speech and non-speech studies.

An equally pressing avenue for further research lies in unraveling the interactions among different training factors. Empirical evidence indicates that training variables rarely operate in isolation; rather, the benefits of variability may depend on the nature of other training factors (Kaipa, 2016; Lavan et al., 2019; Likourezos et al., 2019). Consequently, a promising line of inquiry is to pinpoint how the impact of one variable (e.g., high talker variability) may be amplified or attenuated by another (e.g., explicit feedback or training duration). Systematically characterizing these contextual dependencies will not only refine our theoretical understanding of auditory category learning but also guide the design of optimal, personalized training protocols suited to diverse learner needs and contexts.

## 5. Conclusion

By examining 12 widely studied training variables through multiple complementary analyses and categorizing them into three hidden factors (intensity, variability, and engagement), this study provides new evidence on how these variables influence auditory category learning success. We found that training duration and number of trials (intensity), along with talker variability, context variability, and exemplar variability (variability), plus feedback (engagement), emerged as robust predictors of behavioral improvement.

These findings highlight intensity and variability as crucial drivers of successful auditory category learning. Specifically, immersive, high-exposure protocols and diverse auditory contexts, reinforced by repetitive presentations, may lead to significant improvements in learners' ability to differentiate and internalize new categories. Additionally, our neuroimaging meta-analyses revealed a left-lateralized frontotemporal-parietal network that becomes more active in speech- and motor-related regions, while memory-oriented areas (e.g., cuneus, precuneus) show decreased engagement as learning advances. This newly identified pattern provides deeper insight into the neural outcomes of auditory category learning, suggesting a transition from memory-based strategies to more efficient, speech-motor-driven processing. These findings not only advance theoretical perspectives on auditory category learning but also offer practical guidelines for designing and optimizing future training interventions.

**Data availability statement**

The data and codes supporting the conclusions of this article are available publicly (https://github.com/NanWang0221/acl/tree/main).

**Declaration of generative AI and AI-assisted technologies in the writing process:**

During the preparation of this work, the authors used ChatGPT and Grammarly to polish the writing. After using this tool, they reviewed and edited the content as needed and take full responsibility for the content of the published article.

## References

Acar, F., Seurinck, R., Eickhoff, S. B., & Moerkerke, B. (2018). Assessing robustness against potential publication bias in Activation Likelihood Estimation (ALE) meta-analyses for fMRI. *PLOS ONE*, *13*(11), e0208177. https://doi.org/10.1371/journal.pone.0208177

Anderson, S., Parbery-Clark, A., White-Schwoch, T., & Kraus, N. (2012). Aging affects neural precision of speech encoding. *The Journal of Neuroscience: The Official Journal of the Society for Neuroscience*, *32*(41), 14156–14164. https://doi.org/10.1523/JNEUROSCI.2176-12.2012

Ashby, F. G., & Maddox, W. T. (2005). Human Category Learning. *Annual Review of Psychology*, *56*(Volume 56, 2005), 149–178. https://doi.org/10.1146/annurev.psych.56.091103.070217

Balduzzi, S., Rücker, G., & Schwarzer, G. (2019). How to perform a meta-analysis with R: A practical tutorial. *Evidence-Based Mental Health*, *22*(4), 153–160. https://doi.org/10.1136/ebmental-2019-300117

Bartolotti, J., Bradley, K., Hernandez, A. E., & Marian, V. (2017). Neural signatures of second language learning and control. *NEUROPSYCHOLOGIA*, *98*, 130–138. https://doi.org/10.1016/j.neuropsychologia.2016.04.007

Baumeister, S., Wolf, I., Hohmann, S., Holz, N., Boecker-Schlier, R., Banaschewski, T., & Brandeis, D. (2019). The impact of successful learning of self-regulation on reward processing in children with ADHD using fMRI. *ADHD Attention Deficit and Hyperactivity Disorders*, *11*(1), 31–45. https://doi.org/10.1007/s12402-018-0269-6

Benjamini, Y., & Hochberg, Y. (1995). Controlling the False Discovery Rate: A Practical and Powerful Approach to Multiple Testing. *Journal of the Royal Statistical Society Series B: Statistical Methodology*, *57*(1), 289–300. https://doi.org/10.1111/j.2517-6161.1995.tb02031.x

Ben-Shachar, M. S., Lüdecke, D., & Makowski, D. (2020). effectsize: Estimation of Effect Size Indices and Standardized Parameters. *Journal of Open Source Software*, *5*(56), 2815. https://doi.org/10.21105/joss.02815

Braithwaite, D. W., & Goldstone, R. L. (2015). Effects of Variation and Prior Knowledge on Abstract Concept Learning. *Cognition and Instruction*. https://www.tandfonline.com/doi/abs/10.1080/07370008.2015.1067215

Brekelmans, G., Lavan, N., Saito, H., Clayards, M., & Wonnacott, E. (2022). Does high variability training improve the learning of non-native phoneme contrasts over low variability training? A replication. *Journal of Memory and Language*, *126*, 104352. https://doi.org/10.1016/j.jml.2022.104352

Callan, D. E., Tajima, K., Callan, A. M., Kubo, R., Masaki, S., & Akahane-Yamada, R. (2003). Learning-induced neural plasticity associated with improved identification performance after training of a difficult second-language phonetic contrast. *NeuroImage*, *19*(1), 113–124. https://doi.org/10.1016/s1053-8119(03)00020-x

Carvalho, P. F., & Goldstone, R. L. (2014). Putting category learning in order: Category structure and temporal arrangement affect the benefit of interleaved over blocked study. *Memory & Cognition*, *42*(3), 481–495. https://doi.org/10.3758/s13421-013-0371-0

Carvalho, P. F., & Goldstone, R. L. (2015). What you learn is more than what you see: What can sequencing effects tell us about inductive category learning? *Frontiers in Psychology*, *6*. https://doi.org/10.3389/fpsyg.2015.00505

Cavanna, A. E., & Trimble, M. R. (2006). The precuneus: A review of its functional anatomy and behavioural correlates. *Brain: A Journal of Neurology*, *129*(Pt 3), 564–583. https://doi.org/10.1093/brain/awl004

Censor, N., Karni, A., & Sagi, D. (2006). A link between perceptual learning, adaptation and sleep. *Vision Research*, *46*(23), 4071–4074. https://doi.org/10.1016/j.visres.2006.07.022

Chandrasekaran, B., Yi, H.-G., & Maddox, W. T. (2014). Dual-learning systems during speech category learning. *Psychonomic Bulletin & Review*, *21*(2), 488–495. https://doi.org/10.3758/s13423-013-0501-5

Cox, J., & Witten, I. B. (2019). Striatal circuits for reward learning and decision-making. *Nature Reviews Neuroscience*, *20*(8), 482–494. https://doi.org/10.1038/s41583-019-0189-2

Cramér, H. (1999). *Mathematical methods of statistics* (Vol. 9). Princeton university press. https://books.google.com/books?hl=en&lr=&id=jV2YDwAAQBAJ&oi=fnd&pg=PA3&dq=mathematical+methods+of+statistics&ots=OZvXyfOrAh&sig=nkvvdK-1W3wGfmnvP9ht09qqnnQ

D. Roth. (2005). A latent consolidation phase in auditory identification learning: Time in the awake state is sufficient. *Learning & Memory*, *12 2*, 159–164. https://doi.org/10.1101/87505

Daselaar, S. M., Prince, S. E., Dennis, N. A., Hayes, S. M., Kim, H., & Cabeza, R. (2009). Posterior midline and ventral parietal activity is associated with retrieval success and encoding failure. *Frontiers in Human Neuroscience*, *3*. https://doi.org/10.3389/neuro.09.013.2009

De Souza, A. C. S., Yehia, H. C., Sato, M., & Callan, D. (2013). Brain activity underlying auditory perceptual learning during short period training: Simultaneous fMRI and EEG recording. *BMC Neuroscience*, *14*(1), 8. https://doi.org/10.1186/1471-2202-14-8

Deci, E. L., Koestner, R., & Ryan, R. M. (1999). A meta-analytic review of experiments examining the effects of extrinsic rewards on intrinsic motivation. *Psychological Bulletin*, *125*(6), 627–668. https://doi.org/10.1037/0033-2909.125.6.627

Deng, Y., Chou, T. L., Ding, G. S., Peng, D. L., & Booth, J. R. (2011). The Involvement of Occipital and Inferior Frontal Cortex in the Phonological Learning of Chinese Characters. *JOURNAL OF COGNITIVE NEUROSCIENCE*, *23*(8), 1998–2012. https://doi.org/10.1162/jocn.2010.21571

Desai, R., Liebenthal, E., Waldron, E., & Binder, J. R. (2008). Left Posterior Temporal Regions are Sensitive to Auditory Categorization. *Journal of Cognitive Neuroscience*, *20*(7), 1174–1188. https://doi.org/10.1162/jocn.2008.20081

Dietz, A., Vannest, J., Maloney, T., Altaye, M., Szaflarski, J. P., & Holland, S. K. (2016). The Calculation of Language Lateralization Indices in Post-stroke Aphasia: A Comparison of a Standard and a Lesion-Adjusted Formula. *Frontiers in Human Neuroscience*, *10*. https://www.frontiersin.org/articles/10.3389/fnhum.2016.00493

Doeller, C. F., Opitz, B., Krick, C. M., Mecklinger, A., & Reith, W. (2006). Differential hippocampal and prefrontal-striatal contributions to instance-based and rule-based learning. *NeuroImage*, *31*(4), 1802–1816. https://doi.org/10.1016/j.neuroimage.2006.02.006

Drullman, R., & Bronkhorst, A. W. (2004). Speech perception and talker segregation: Effects of level, pitch, and tactile support with multiple simultaneous talkers. *The Journal of the Acoustical Society of America*, *116*(5), 3090–3098. https://doi.org/10.1121/1.1802535

Eickhoff, S. B., Bzdok, D., Laird, A. R., Kurth, F., & Fox, P. T. (2012). Activation likelihood estimation meta-analysis revisited. *NeuroImage*, *59*(3), 2349–2361. https://doi.org/10.1016/j.neuroimage.2011.09.017

Eickhoff, S. B., Laird, A. R., Grefkes, C., Wang, L. E., Zilles, K., & Fox, P. T. (2009). Coordinate-based activation likelihood estimation meta-analysis of neuroimaging data: A random-effects approach based on empirical estimates of spatial uncertainty. *Human Brain Mapping*, *30*(9), 2907–2926. https://doi.org/10.1002/hbm.20718

Enge, A., Abdel Rahman, R., & Skeide, M. A. (2021). A meta-analysis of fMRI studies of semantic cognition in children. *NeuroImage*, *241*, 118436. https://doi.org/10.1016/j.neuroimage.2021.118436

Estes, K. G., & Lew-Williams, C. (2015). Listening through voices: Infant statistical word segmentation across multiple speakers. *Developmental Psychology*, *51*(11), 1517–1528. https://doi.org/10.1037/a0039725

Estes, W. K., & Burke, C. J. (1953). A theory of stimulus variability in learning. *Psychological Review*, *60*(4), 276–286. https://doi.org/10.1037/h0055775

Feng, G., Gan, Z., Yi, H. G., Ell, S. W., Roark, C. L., Wang, S., Wong, P. C. M., & Chandrasekaran, B. (2021). Neural dynamics underlying the acquisition of distinct auditory category structures. *NeuroImage*, *244*, 118565. https://doi.org/10.1016/j.neuroimage.2021.118565

Feng, G., Yi, H. G., & Chandrasekaran, B. (2019). The Role of the Human Auditory Corticostriatal Network in Speech Learning. *Cerebral Cortex*, *29*(10), 4077–4089. https://doi.org/10.1093/cercor/bhy289

Filoteo, J. V., Maddox, W. T., Simmons, A. N., Ing, A. D., Cagigas, X. E., Matthews, S., & Paulus, M. P. (2005). Cortical and subcortical brain regions involved in rule-based category learning. *NeuroReport*, *16*(2), 111.

Fredricks, J. A., Blumenfeld, P. C., & Paris, A. H. (2004). School Engagement: Potential of the Concept, State of the Evidence. *Review of Educational Research*, *74*(1), 59–109. https://doi.org/10.3102/00346543074001059

Friederici, A. D., & Alter, K. (2004). Lateralization of auditory language functions: A dynamic dual pathway model. *Brain and Language*, *89*(2), 267–276. https://doi.org/10.1016/S0093-934X(03)00351-1

Gabay, Y., Dick, F. K., Zevin, J. D., & Holt, L. L. (2015). Incidental auditory category learning. *Journal of Experimental Psychology: Human Perception and Performance*, *41*(4), 1124–1138. https://doi.org/10.1037/xhp0000073

Galvan, A., Hare, T. A., Davidson, M., Spicer, J., Glover, G., & Casey, B. J. (2005). The Role of Ventral Frontostriatal Circuitry in Reward-Based Learning in Humans. *Journal of Neuroscience*, *25*(38), 8650–8656. https://doi.org/10.1523/JNEUROSCI.2431-05.2005

Gan, Z., Zheng, L., Wang, S., & Feng, G. (2023). Distribution-dependent representations in auditory category learning and generalization. *Frontiers in Psychology*, *14*. https://www.frontiersin.org/journals/psychology/articles/10.3389/fpsyg.2023.1132570

Goh, W. D. (2005). Talker variability and recognition memory: Instance-specific and voice-specific effects. *Journal of Experimental Psychology. Learning, Memory, and Cognition*, *31*(1), 40–53. https://doi.org/10.1037/0278-7393.31.1.40

Goldstone, R. L., & Hendrickson, A. T. (2010). Categorical perception. *Wiley Interdisciplinary Reviews. Cognitive Science*, *1*(1), 69–78. https://doi.org/10.1002/wcs.26

Gómez, R. L. (2002). Variability and Detection of Invariant Structure. *Psychological Science*, *13*(5), 431–436. https://doi.org/10.1111/1467-9280.00476

Grady, C. L., Rieck, J. R., Nichol, D., Rodrigue, K. M., & Kennedy, K. M. (2020). Influence of sample size and analytic approach on stability and interpretation of brain-behavior correlations in task-related fMRI data. *Human Brain Mapping*, *42*(1), 204–219. https://doi.org/10.1002/hbm.25217

Grinband, J., Hirsch, J., & Ferrera, V. (2006). A Neural Representation of Categorization Uncertainty in the Human Brain. *Neuron*, *49*, 757–763. https://doi.org/10.1016/j.neuron.2006.01.032

Guo, Z., Chen, J., Liu, S., Li, Y., Sun, B., & Gao, Z. (2013). Brain areas activated by uncertain reward-based decision-making in healthy volunteers. *Neural Regeneration Research*, *8*(35), 3344. https://doi.org/10.3969/j.issn.1673-5374.2013.35.009

Hammill, J., Nguyen, T., & Henderson, F. (2022). Student engagement: The impact of positive psychology interventions on students. *Active Learning in Higher Education*, *23*(2), 129–142. https://doi.org/10.1177/1469787420950589

Hardison, D. M. (2003). Acquisition of second-language speech: Effects of visual cues, context, and talker variability. *Applied Psycholinguistics*, *24*(4), 495–522. https://doi.org/10.1017/S0142716403000250

Hattie, J., & Timperley, H. (2007). The Power of Feedback. *Review of Educational Research*, *77*(1), 81–112. https://doi.org/10.3102/003465430298487

Hedges, L., & Olkin, I. (1985). Statistical Methods in Meta-Analysis. In *Stat Med* (Vol. 20). https://doi.org/10.2307/1164953

Helie, S., Roeder, J. L., & Ashby, F. G. (2010). Evidence for Cortical Automaticity in Rule-Based Categorization. *Journal of Neuroscience*, *30*(42), 14225–14234. https://doi.org/10.1523/JNEUROSCI.2393-10.2010

Heo, S., Sung, Y., & Lee, S. W. (2021). Effects of subclinical depression on prefrontal–striatal model-based and model-free learning. *PLOS Computational Biology*, *17*(5), e1009003. https://doi.org/10.1371/journal.pcbi.1009003

Hickok, G., & Poeppel, D. (2000). Towards a functional neuroanatomy of speech perception. *Trends in Cognitive Sciences*, *4*(4), 131–138. https://doi.org/10.1016/s1364-6613(00)01463-7

Hickok, G., & Poeppel, D. (2007). The cortical organization of speech processing. *Nature Reviews Neuroscience*, *8*(5), Article 5. https://doi.org/10.1038/nrn2113

Higgins, J. P. T., Thompson, S. G., Deeks, J. J., & Altman, D. G. (2003). Measuring inconsistency in meta-analyses. *BMJ*, *327*(7414), 557–560. https://doi.org/10.1136/bmj.327.7414.557

Holt, L. L., & Lotto, A. J. (2010). Speech perception as categorization. *Attention, Perception, & Psychophysics*, *72*(5), 1218–1227. https://doi.org/10.3758/APP.72.5.1218

James, T. W., & Gauthier, I. (2003). Auditory and Action Semantic Features Activate Sensory-Specific Perceptual Brain Regions. *Current Biology*, *13*(20), 1792–1796. https://doi.org/10.1016/j.cub.2003.09.039

Jochen Kaiser, & Werner Lutzenberger. (2005). Cortical Oscillatory Activity and the Dynamics of Auditory Memory Processing. *Reviews in the Neurosciences*, *16*(3), 239–254. https://doi.org/10.1515/REVNEURO.2005.16.3.239

Joiner, W. M., & Smith, M. A. (2008). Long-term retention explained by a model of short-term learning in the adaptive control of reaching. *Journal of Neurophysiology*, *100 5*, 2948–2955. https://doi.org/10.1152/jn.90706.2008

Kaipa, R. (2016). Is There an Interaction between Task Complexity and Practice Variability in Speech-Motor learning? *Annals of Neurosciences*, *23*(3), 134–138. https://doi.org/10.1159/000449178

Kartushina, N., & Martin, C. D. (2019). Talker and Acoustic Variability in Learning to Produce Nonnative Sounds: Evidence from Articulatory Training. *Language Learning*, *69*(1), 71–105. https://doi.org/10.1111/lang.12315

Karuza, E. A., Emberson, L. L., & Aslin, R. N. (2014). Combining fMRI and behavioral measures to examine the process of human learning. *Neurobiology of Learning and Memory*, *109*, 193–206. https://doi.org/10.1016/j.nlm.2013.09.012

Kaufman, S. B., Deyoung, C. G., Gray, J. R., Jiménez, L., Brown, J., & Mackintosh, N. (2010). Implicit learning as an ability. *Cognition*, *116*(3), 321–340. https://doi.org/10.1016/j.cognition.2010.05.011

Kirschen, M. P., Chen, S. H. A., Schraedley-Desmond, P., & Desmond, J. E. (2005). Load- and practice-dependent increases in cerebro-cerebellar activation in verbal working memory: An fMRI study. *NeuroImage*, *24*(2), 462–472. https://doi.org/10.1016/j.neuroimage.2004.08.036

Kleim, J. A., & Jones, T. A. (2008). Principles of experience-dependent neural plasticity: Implications for rehabilitation after brain damage. *Journal of Speech, Language, and Hearing Research: JSLHR*, *51*(1), S225-239. https://doi.org/10.1044/1092-4388(2008/018)

Kobayashi, K., Shiba, Y., Honda, S., Nakajima, S., Fujii, S., Mimura, M., & Noda, Y. (2024). Short-Term Effect of Auditory Stimulation on Neural Activities: A Scoping Review of Longitudinal Electroencephalography and Magnetoencephalography Studies. *Brain Sciences*, *14*(2), Article 2. https://doi.org/10.3390/brainsci14020131

Krishnan, S., Watkins, K., & Bishop, D. (2016). Neurobiological Basis of Language Learning Difficulties. *Trends in Cognitive Sciences*, *xx*. https://doi.org/10.1016/j.tics.2016.06.012

Kuhl, P. K. (2004a). Early language acquisition: Cracking the speech code. *Nature Reviews. Neuroscience*, *5*(11), 831–843. https://doi.org/10.1038/nrn1533

Kuhl, P. K. (2004b). Early language acquisition: Cracking the speech code. *Nature Reviews. Neuroscience*, *5*(11), 831–843. https://doi.org/10.1038/nrn1533

Kuhl, P. K., Stevens, E., Hayashi, A., Deguchi, T., Kiritani, S., & Iverson, P. (2006). Infants show a facilitation effect for native language phonetic perception between 6 and 12 months. *Developmental Science*, *9*(2), F13–F21. https://doi.org/10.1111/j.1467-7687.2006.00468.x

Lammert, A., Proctor, M., & Narayanan, S. (2013). Morphological variation in the adult hard palate and posterior pharyngeal wall. *Journal of Speech, Language, and Hearing Research: JSLHR*, *56*(2), 521–530. https://doi.org/10.1044/1092-4388(2012/12-0059)

Lavan, N., Knight, S., Hazan, V., & McGettigan, C. (2019). The effects of high variability training on voice identity learning. *The Journal of the Acoustical Society of America*, *146*(4_Supplement), 3053–3054. https://doi.org/10.1121/1.5137589

Liang, B., & Du, Y. (2018). The Functional Neuroanatomy of Lexical Tone Perception: An Activation Likelihood Estimation Meta-Analysis. *Frontiers in Neuroscience*, *12*. https://doi.org/10.3389/fnins.2018.00495

Likourezos, V., Kalyuga, S., & Sweller, J. (2019). The Variability Effect: When Instructional Variability Is Advantageous. *Educational Psychology Review*, *31*(2), 479–497. https://doi.org/10.1007/s10648-019-09462-8

Lim, S., & Holt, L. L. (2011). Learning foreign sounds in an alien world: Videogame training improves non-native speech categorization. *Cognitive Science*, *35*(7), 1390–1405. https://doi.org/10.1111/j.1551-6709.2011.01192.x

Lim, S.-J., Fiez, J. A., & Holt, L. L. (2014). How may the basal ganglia contribute to auditory categorization and speech perception? *Frontiers in Neuroscience*, *8*. https://doi.org/10.3389/fnins.2014.00230

Lim, S.-J., Fiez, J. A., & Holt, L. L. (2019). Role of the striatum in incidental learning of sound categories. *Proceedings of the National Academy of Sciences*, *116*(10), 4671–4680. https://doi.org/10.1073/pnas.1811992116

Lipsey, M. W., & Wilson, D. (2000). *Practical Meta-Analysis* (1st edition). SAGE Publications, Inc.

Lively, S. E., Logan, J. S., & Pisoni, D. B. (1993). Training Japanese listeners to identify English /r/ and /l/. II: The role of phonetic environment and talker variability in learning new perceptual categories. *The Journal of the Acoustical Society of America*, *94*(3), 1242–1255. https://doi.org/10.1121/1.408177

Logan, J. S., Lively, S. E., & Pisoni, D. B. (1991). Training Japanese listeners to identify English /r/ and /l/: A first report. *The Journal of the Acoustical Society of America*, *89*(2), 874–886. https://doi.org/10.1121/1.1894649

Lohse, K. R., Wadden, K., Boyd, L. A., & Hodges, N. J. (2014). Motor skill acquisition across short and long time scales: A meta-analysis of neuroimaging data. *Neuropsychologia*, *59*, 130–141. https://doi.org/10.1016/j.neuropsychologia.2014.05.001

López-Barroso, D., & de Diego-Balaguer, R. (2017). Language Learning Variability within the Dorsal and Ventral Streams as a Cue for Compensatory Mechanisms in Aphasia Recovery. *Frontiers in Human Neuroscience*, *11*. https://doi.org/10.3389/fnhum.2017.00476

Maddox, W. T., Chandrasekaran, B., Smayda, K., & Yi, H.-G. (2013). Dual systems of speech category learning across the lifespan. *Psychology and Aging*, *28*(4), 1042–1056. https://doi.org/10.1037/a0034969

Maddox, W. T., Love, B. C., Glass, B. D., & Filoteo, J. V. (2008a). When more is less: Feedback effects in perceptual category learning. *Cognition*, *108*(2), 578–589. https://doi.org/10.1016/j.cognition.2008.03.010

Maddox, W. T., Love, B. C., Glass, B. D., & Filoteo, J. V. (2008b). When more is less: Feedback effects in perceptual category learning. *Cognition*, *108*(2), 578–589. https://doi.org/10.1016/j.cognition.2008.03.010

McCandliss, B. D., Fiez, J. A., Protopapas, A., Conway, M., & McClelland, J. L. (2002). Success and failure in teaching the [r]-[l] contrast to Japanese adults: Tests of a Hebbian model of plasticity and stabilization in spoken language perception. *Cognitive, Affective & Behavioral Neuroscience*, *2*(2), 89–108. https://doi.org/10.3758/cabn.2.2.89

McMurray, B. (2023). The acquisition of speech categories: Beyond perceptual narrowing, beyond unsupervised learning and beyond infancy. *Language, Cognition and Neuroscience*, *38*(4), 419–445. https://doi.org/10.1080/23273798.2022.2105367

McMurray, B., & Aslin, R. N. (2005). Infants are sensitive to within-category variation in speech perception. *Cognition*, *95*(2), B15–B26. https://doi.org/10.1016/j.cognition.2004.07.005

Molloy, K., Moore, D. R., Sohoglu, E., & Amitay, S. (2012). Less Is More: Latent Learning Is Maximized by Shorter Training Sessions in Auditory Perceptual Learning. *PLOS ONE*, *7*(5), e36929. https://doi.org/10.1371/journal.pone.0036929

Müller, V., Cieslik, E., Laird, A., Fox, P., Radua, J., Mataix-Cols, D., Tench, C., Yarkoni, T., Nichols, T., Turkeltaub, P., Wager, T., & Eickhoff, S. (2017). Ten simple rules for neuroimaging meta-analysis. *Neuroscience & Biobehavioral Reviews*, *84*. https://doi.org/10.1016/j.neubiorev.2017.11.012

Myers, E. B. (2014). Emergence of category-level sensitivities in non-native speech sound learning. *Frontiers in Neuroscience*, *8*. https://doi.org/10.3389/fnins.2014.00238

Myers, E. B., & Swan, K. (2012). Effects of Category Learning on Neural Sensitivity to Non-native Phonetic Categories. *Journal of Cognitive Neuroscience*, *24*(8), 1695–1708. https://doi.org/10.1162/jocn_a_00243

Newman-Norlund, R. D., Frey, S. H., Petitto, L.-A., & Grafton, S. T. (2006). Anatomical Substrates of Visual and Auditory Miniature Second-language Learning. *Journal of Cognitive Neuroscience*, *18*(12), 1984–1997. https://doi.org/10.1162/jocn.2006.18.12.1984

Nishi, K., & Kewley, -Port Diane. (2007). Training Japanese Listeners to Perceive American English Vowels: Influence of Training Sets. *Journal of Speech, Language, and Hearing Research*, *50*(6), 1496–1509. https://doi.org/10.1044/1092-4388(2007/103)

Obasih, C. O., Luthra, S., Dick, F., & Holt, L. L. (2023). Auditory category learning is robust across training regimes. *Cognition*, *237*, 105467. https://doi.org/10.1016/j.cognition.2023.105467

Ohl, F., Scheich, H., & Freeman, W. (2001). Change in pattern of ongoing cortical activity with auditory category learning. *Nature*, *412*, 733–736. https://doi.org/10.1038/35089076

Olejnik, S., & Algina, J. (2003). Generalized Eta and Omega Squared Statistics: Measures of Effect Size for Some Common Research Designs. *Psychological Methods*, *8*(4), 434–447. https://doi.org/10.1037/1082-989X.8.4.434

Orwin, R. G. (1983). A fail-safe N for effect size in meta-analysis. *Journal of Educational Statistics*, *8*(2), 157–159. https://doi.org/10.2307/1164923

Pedrosa, V., Menichini, E., Pajot-Moric, Q., Vincent, P., Zhou, L., Teachen, L., Latham, P., & Akrami, A. (2023). *Humans, rats and mice show species-specific adaptations to sensory statistics in categorisation behaviour* (p. 2023.01.30.526119). bioRxiv. https://doi.org/10.1101/2023.01.30.526119

Perrachione, T. K., Lee, J., Ha, L. Y. Y., & Wong, P. C. M. (2011). Learning a novel phonological contrast depends on interactions between individual differences and training paradigm design. *The Journal of the Acoustical Society of America*, *130*(1), 461–472. https://doi.org/10.1121/1.3593366

Plihal, W., & Born, J. (1997). Effects of early and late nocturnal sleep on declarative and procedural memory. *Journal of Cognitive Neuroscience*, *9*(4), 534–547. https://doi.org/10.1162/jocn.1997.9.4.534

Poeppel, D., Guillemin, A., Thompson, J., Fritz, J., Bavelier, D., & Braun, A. R. (2004). Auditory lexical decision, categorical perception, and FM direction discrimination differentially engage left and right auditory cortex. *Neuropsychologia*, *42*(2), 183–200. https://doi.org/10.1016/j.neuropsychologia.2003.07.010

Poldrack, R. A. (2011). Inferring Mental States from Neuroimaging Data: From Reverse Inference to Large-Scale Decoding. *Neuron*, *72*(5), 692–697. https://doi.org/10.1016/j.neuron.2011.11.001

Poldrack, R. A., Baker, C. I., Durnez, J., Gorgolewski, K. J., Matthews, P. M., Munafò, M. R., Nichols, T. E., Poline, J.-B., Vul, E., & Yarkoni, T. (2017). Scanning the horizon: Towards transparent and reproducible neuroimaging research. *Nature Reviews. Neuroscience*, *18*(2), 115–126. https://doi.org/10.1038/nrn.2016.167

Pulvermüller, F., Huss, M., Kherif, F., Moscoso del Prado Martin, F., Hauk, O., & Shtyrov, Y. (2006). Motor cortex maps articulatory features of speech sounds. *Proceedings of the National Academy of Sciences*, *103*(20), 7865–7870. https://doi.org/10.1073/pnas.0509989103

Raviv, L., Lupyan, G., & Green, S. C. (2022). How variability shapes learning and generalization. *Trends in Cognitive Sciences*, *26*(6), 462–483. https://doi.org/10.1016/j.tics.2022.03.007

Roark, C. L., & Chandrasekaran, B. (2023). Stable, flexible, common, and distinct behaviors support rule-based and information-integration category learning. *NPJ Science of Learning*, *8*(1), 14. https://doi.org/10.1038/s41539-023-00163-0

Roark, C. L., & Holt, L. L. (2019). Perceptual dimensions influence auditory category learning. *Attention, Perception, & Psychophysics*, *81*(4), 912–926. https://doi.org/10.3758/s13414-019-01688-6

Rohrer, D., & Taylor, K. (2006). The effects of overlearning and distributed practise on the retention of mathematics knowledge. *Applied Cognitive Psychology*, *20*(9), 1209–1224. https://doi.org/10.1002/acp.1266

Roux, F., & Uhlhaas, P. J. (2014). Working memory and neural oscillations: Alpha–gamma versus theta–gamma codes for distinct WM information? *Trends in Cognitive Sciences*, *18*(1), 16–25. https://doi.org/10.1016/j.tics.2013.10.010

Sadakata, M., & McQueen, J. M. (2013). High stimulus variability in nonnative speech learning supports formation of abstract categories: Evidence from Japanese geminates. *The Journal*

*of the Acoustical Society of America*, *134*(2), 1324–1335. https://doi.org/10.1121/1.4812767

Saffran, J. R., Aslin, R. N., & Newport, E. L. (1996). Statistical learning by 8-month-old infants. *Science (New York, N.Y.)*, *274*(5294), 1926–1928. https://doi.org/10.1126/science.274.5294.1926

Samartsidis, P., Montagna, S., Laird, A. R., Fox, P. T., Johnson, T. D., & Nichols, T. E. (2020). Estimating the prevalence of missing experiments in a neuroimaging meta-analysis. *Research Synthesis Methods*, *11*(6), 866–883. https://doi.org/10.1002/jrsm.1448

Saur, D., Kreher, B. W., Schnell, S., Kümmerer, D., Kellmeyer, P., Vry, M.-S., Umarova, R., Musso, M., Glauche, V., Abel, S., Huber, W., Rijntjes, M., Hennig, J., & Weiller, C. (2008). Ventral and dorsal pathways for language. *Proceedings of the National Academy of Sciences of the United States of America*, *105*(46), 18035–18040. https://doi.org/10.1073/pnas.0805234105

Schönberg, T., Daw, N. D., Joel, D., & O'Doherty, J. P. (2007). Reinforcement Learning Signals in the Human Striatum Distinguish Learners from Nonlearners during Reward-Based Decision Making. *Journal of Neuroscience*, *27*(47), 12860–12867. https://doi.org/10.1523/JNEUROSCI.2496-07.2007

Schorn, J. M., & Knowlton, B. J. (2021). Interleaved practice benefits implicit sequence learning and transfer. *Memory & Cognition*, *49*(7), 1436–1452. https://doi.org/10.3758/s13421-021-01168-z

Shi, L., & Lin, L. (2019). The trim-and-fill method for publication bias: Practical guidelines and recommendations based on a large database of meta-analyses. *Medicine*, *98*(23), e15987. https://doi.org/10.1097/MD.0000000000015987

Sinkeviciute, R., Brown, H., Brekelmans, G., & Wonnacott, E. (2019). THE ROLE OF INPUT VARIABILITY AND LEARNER AGE IN SECOND LANGUAGE VOCABULARY LEARNING. *Studies in Second Language Acquisition*, *41*(4), 795–820. https://doi.org/10.1017/S0272263119000263

Smith, E. E., & Grossman, M. (2008). Multiple Systems of Category Learning. *Neuroscience and Biobehavioral Reviews*, *32*(2), 249–264. https://doi.org/10.1016/j.neubiorev.2007.07.009

Smith, J. D., Berg, M., Cook, R. G., Murphy, M. S., Crossley, M., Boomer, J., Spiering, B. J., Beran, M., Church, B., Ashby, F. G., Grace, R., Edu, M., Edu, & Smith, D. (2012). Implicit and explicit categorization: A tale of four species. *Neuroscience & Biobehavioral Reviews*, *36*, 2355–2369. https://doi.org/10.1016/j.neubiorev.2012.09.003

Stein, M., Federspiel, A., Koenig, T., Wirth, M., Lehmann, C., Wiest, R., Strik, W., Brandeis, D., & Dierks, T. (2009). Reduced frontal activation with increasing 2nd language proficiency. *Neuropsychologia*, *47*(13), 2712–2720. https://doi.org/10.1016/j.neuropsychologia.2009.05.023

Studdert-Kennedy, M., & Shankweiler, D. (1970). Hemispheric specialization for speech perception. *The Journal of the Acoustical Society of America*, *48*(2), 579–594. https://doi.org/10.1121/1.1912174

Szaflarski, J. P., Holland, S. K., Schmithorst, V. J., & Byars, A. W. (2005). fMRI study of language lateralization in children and adults. *Human Brain Mapping*, *27*(3), 202–212. https://doi.org/10.1002/hbm.20177

Tang, H., Crain, S., & Johnson, B. W. (2016). Dual temporal encoding mechanisms in human auditory cortex: Evidence from MEG and EEG. *NeuroImage*, *128*, 32–43. https://doi.org/10.1016/j.neuroimage.2015.12.053

Tervaniemi, M., & Hugdahl, K. (2003). Lateralization of auditory-cortex functions. *Brain Research. Brain Research Reviews*, *43*(3), 231–246. https://doi.org/10.1016/j.brainresrev.2003.08.004

Thorin, J., Sadakata, M., Desain, P., & McQueen, J. M. (2018). Perception and production in interaction during non-native speech category learning. *The Journal of the Acoustical Society of America*, *144*(1), 92. https://doi.org/10.1121/1.5044415

Trowler, V. (2010). Student engagement literature review. *The Higher Education Academy*, *11*(1), 1–15.

Tucker, M. A., Hirota, Y., Wamsley, E. J., Lau, H., Chaklader, A., & Fishbein, W. (2006). A daytime nap containing solely non-REM sleep enhances declarative but not procedural memory. *Neurobiology of Learning and Memory*, *86*(2), 241–247. https://doi.org/10.1016/j.nlm.2006.03.005

Turkeltaub, P. E., Eden, G. F., Jones, K. M., & Zeffiro, T. A. (2002). Meta-analysis of the functional neuroanatomy of single-word reading: Method and validation. *NeuroImage*, *16*(3 Pt 1), 765–780. https://doi.org/10.1006/nimg.2002.1131

Viechtbauer, W. (2010). Conducting Meta-Analyses in R with the metafor Package. *Journal of Statistical Software*, *36*, 1–48. https://doi.org/10.18637/jss.v036.i03

Vlahou, E., Protopapas, A., & Seitz, A. (2011). Implicit Training of Nonnative Speech Stimuli. *Journal of Experimental Psychology. General*, *141*, 363–381. https://doi.org/10.1037/a0025014

Wang, Y., Sereno, J. A., Jongman, A., & Hirsch, J. (2003a). fMRI evidence for cortical modification during learning of Mandarin lexical tone. *JOURNAL OF COGNITIVE NEUROSCIENCE*, *15*(7), 1019–1027. https://doi.org/10.1162/089892903770007407

Wang, Y., Sereno, J. A., Jongman, A., & Hirsch, J. (2003b). fMRI Evidence for Cortical Modification during Learning of Mandarin Lexical Tone. *Journal of Cognitive Neuroscience*, *15*(7), 1019–1027. https://doi.org/10.1162/089892903770007407

Watanabe, T., & Sasaki, Y. (2015). Perceptual Learning: Toward a Comprehensive Theory. *Annual Review of Psychology*, *66*(Volume 66, 2015), 197–221. https://doi.org/10.1146/annurev-psych-010814-015214

Werker, J. F., & Hensch, T. K. (2015). Critical periods in speech perception: New directions. *Annual Review of Psychology*, *66*, 173–196. https://doi.org/10.1146/annurev-psych-010814-015104

Wong, P. C. M., Jin, J. X., Gunasekera, G. M., Abel, R., Lee, E. R., & Dhar, S. (2009). Aging and cortical mechanisms of speech perception in noise. *Neuropsychologia*, *47*(3), 693–703. https://doi.org/10.1016/j.neuropsychologia.2008.11.032

Wong, P. C. M., Nusbaum, H. C., & Small, S. L. (2004). Neural bases of talker normalization. *Journal of Cognitive Neuroscience*, *16*(7), 1173–1184. https://doi.org/10.1162/0898929041920522

Worthy, D. A., Markman, A. B., & Todd Maddox, W. (2013). Feedback and stimulus-offset timing effects in perceptual category learning. *Brain and Cognition*, *81*(2), 283–293. https://doi.org/10.1016/j.bandc.2012.11.006

Xia, M., Wang, J., & He, Y. (2013). BrainNet Viewer: A Network Visualization Tool for Human Brain Connectomics. *PLoS ONE*, *8*(7), e68910. https://doi.org/10.1371/journal.pone.0068910

Yarkoni, T., Poldrack, R. A., Nichols, T. E., Van Essen, D. C., & Wager, T. D. (2011). Large-scale automated synthesis of human functional neuroimaging data. *Nature Methods*, *8*(8), 665–670. https://doi.org/10.1038/nmeth.1635

Yi, H.-G., Maddox, W. T., Mumford, J. A., & Chandrasekaran, B. (2016). The Role of Corticostriatal Systems in Speech Category Learning. *Cerebral Cortex*, *26*(4), 1409–1420. https://doi.org/10.1093/cercor/bhu236

Yoo, S.-S., Lee, J.-H., O'Leary, H., Lee, V., Choo, S.-E., & Jolesz, F. A. (2007). Functional magnetic resonance imaging-mediated learning of increased activity in auditory areas. *NeuroReport*, *18*(18), 1915–1920. https://doi.org/10.1097/WNR.0b013e3282f202ac

Yotsumoto, Y., Watanabe, T., & Sasaki, Y. (2008). Different Dynamics of Performance and Brain Activation in the Time Course of Perceptual Learning. *Neuron*, *57*(6), 827–833. https://doi.org/10.1016/j.neuron.2008.02.034

Zatorre, R. J., Belin, P., & Penhune, V. B. (2002). Structure and function of auditory cortex: Music and speech. *Trends in Cognitive Sciences*, *6*(1), 37–46. https://doi.org/10.1016/s1364-6613(00)01816-7

Zatorre, R. J., Delhommeau, K., & Zarate, J. M. (2012). Modulation of auditory cortex response to pitch variation following training with microtonal melodies. *FRONTIERS IN PSYCHOLOGY*, *3*. https://doi.org/10.3389/fpsyg.2012.00544